\documentstyle[art12]{article}
 \def\centereps#1#2#3{\vskip#2\relax\centerline{
 \hbox to#1{\special{eps:#3 x=#1, y=#2}\hfil}}}

\def\ind#1{\buildrel #1 \over \longrightarrow}
\def\Ind#1{\buildrel #1 \over \Longrightarrow}

\def\D{{\bf D}}      \def\H{{\bf H}} \def\K{{\bf K}}
   \def\T{{\bf T}}

   \def\gd{\delta}  
\def\gs{\sigma}    \def\gD{\Delta}  
\def\gG{\Gamma}   \def\gS{\Delta}  \def\gL{\Lambda} \def\gS{\Sigma}
\def\gY{\Upsilon}   

     \def\G{{\bf G}}    \def\g{{\bf g}}

\def\C{{\bf C}}   
\def\M{{\rm M}}  
\def\X{{\rm X}}  \def\Y{{\rm Y}} 
\def\gr{{\rm gr}}

\def\zhs{(\gs_h|\frac{1}{2}\,\frac{1}{2})}

\def\zvs{(\gs|\frac{1}{2}\,\frac{1}{2})}
\def\zux{(U_x|\frac{1}{2}\,0)}
\def\zus{(U|\frac{1}{2}\,\frac{1}{2})}

\def\tx{(I|1\, 0)}
\def\ty{(I|0\, 1)}

\def\zhy{(\gs_h|0\,\frac{1}{2})}
\def\zvy{(\gs_y|0\,\frac{1}{2})}
\def\zuy{(U_y|0 \frac{1}{2})}
\def\zts{(I|\frac{1}{2} \frac{1}{2})}

                    \def\d={\buildrel \rm def \over =}
 
\def\diag{{\rm diag}\,}

\def\Brillouin{Brillouin\ }

\begin{document}
\baselineskip=0.2in
\begin{titlepage}
\pagestyle{empty}
\title{{\bf Irreducible Representations of Diperiodic Groups}}
\author{}
\author{Ivanka Milo\v sevi\' c$^{\dagger\ast}$, B. Nikoli\'c$^\dagger$,
M. Damnjanovi\'c$^\dagger$, Maja Kr\v cmar$^\ddagger$\\ 
$^\dagger$ Faculty of Physics, The University,\\ POB 368, 11001 Beograd, Yugoslavia\\
$^\ddagger$ Department of Physics, Texas A\& M University\\
College Station, TX 77843-4242, USA}
\maketitle
\vskip3cm
\begin{flushbottom}\begin{abstract}
The irreducible representations of all of the 80 diperiodic groups,
being the symmetries of the systems translationally periodical in two
directions, are calculated. To this end, each of these groups is
factorized as the product of a generalized translational group and an
axial point group.  The results are presented in the form of the
tables, containing the matrices of the irreducible representations of
the generators of the groups. General properties and some physical
applications (degeneracy and topology of the energy bands, selection rules,
etc.) are discussed.
\vskip 1.5cm
$^\ast\,$E-mail: ivag@afrodita.rcub.bg.ac.yu
\end{abstract}\end{flushbottom}
\end{titlepage}
\section{Introduction}
The diperiodic groups, \cite{EW,KOPSKY}, are the symmetry groups of the
systems with translational periodicity in two directions. Thin layers
and multilayers are the obvious examples of such systems.  The interest
for the diperiodic structures has increased after it has been observed
that the $CuO_2$ layers are responsible for the high temperature
superconductivity \cite{PLAC}, and that the main effects (including
superconductivity and the unusual conducting properties above $T_c$)
are present even if there is no periodicity in the direction orthogonal
onto conducting block.

There are 80 diperiodic groups; 17 of them are planar, i. e. describe
the symmetry of the strictly two-dimensional (2D) systems.  All of them
are subgroups of the 3D space groups; this correspondence (involving
additional conventions to reduce the nonuniquenss --- for each space
group a number of diperiodic subgroups can be found, while each
diperiodic group is subgroup in various space groups) has been
determined, \cite{KOPSKY,HA-STO}. The orbits and stabilizers of the
diperiodic groups are known \cite{EW}. In contrast, there is no general
tabulation of the irreducible representations (irrs), beyond the more
specialized tables of Hatch and Stokes \cite{HA-STO} (irrs related to
the points of symmetry in the Brillouin zone are considered in the
context of phase transitions). This may be the reason why the rare
usage of the diperiodic groups in the literature (in contrast to the
space, the point and the line groups) mostly refers to the IC and Raman
spectra, when the irrs of the isogonal point group are effectively
employed. In fact, the most of the results refere to the phase
transitions: Hatch and Stokes found also Molien functions and
invariants\cite{HA-STO}.  The aim of this paper is to fill in this gap,
and to construct the irrs of all the diperiodic groups, thus enabling
extensive treatment of this type of symmetry in the solid state
physics.

In the next section, the specific structural properties of the
diperiodic groups are reviewed. These allow the simple construction of
the irrs, involving neither projective representations, nor the
representations of the space supergroups. The results are given in the
section \ref{MAIN}. Finally, the concluding remarks summarize some
general properties of the irrs in view of their physical applications.
\section{Group structure and construction of irrs}
Each diperiodic group $\D\g$ can be factorized as a weak direct product
of the generalized translational group ${\bf Z}$ and the axial point
group ${\bf P}$: $\D\g ={\bf ZP}$.  This is analogous to the line groups,
\cite{YV}, except that the generalized translational group, ${\bf
Z}$, is two dimensional, since it describes the periodical arrangement
of the elementary motifs along two independent directions (these two
directions are assumed to be in the $xy$-plane). Therefore, ${\bf Z}$
can be formed of the generalized 1D translational groups leaving the
$xy$-plane invariant. There are only four generalized 1D translational
groups satisfying this condition:
\begin{enumerate}
\item pure translational group $T$ along an axis in the plane, 
\item screw axis group $2_1$ with the $C_2$ axis in the plane, 
\item glide plane group $T_h$ of the horizontal, $xy$, glide plane,
\item glide plane group $T_v$ of the vertical glide plane 
(containing $z$ axis).
\end{enumerate}
All these groups are infinite cyclic groups. The first of them is
generated by pure translation; as for the remaining 3 groups, pure
translations are the index-two subgroup generated by the square of
the generator of the glide plane or the screw axis.
The generalized 2D translational groups
are direct or weak direct products of the listed four 1D
generalized translations:
\begin{enumerate}
\item the pure 2D translational group ${\bf T}$ is the direct product of
the two 1D translational groups $T$ along independent directions, with,
in general, different translational periods and an arbitrary angle
between the translational directions.
\item the horizontal 2D glide plane group ${\bf T}_h=TT_h$; 
translational periods of $T$ and $T_h$ may be different, and their
directions form an arbitrary angle.
\item the 2D screw axis group ${\bf 2}_1=T2_1$ (horizontal screw axis);
\item the vertical 2D glide plane group ${\bf T}_v=TT_v$ (vertical glide plane).
\item the product $2_1T_h$ of the groups $2_1$ and $T_h$ generated by 
$(U_x|\frac{1}{2}\,0)$ and $(\gs_h|0\,\frac{1}{2})$.
\end{enumerate}
In the last three cases the screw axis (glide plane) can be chosen in
the direction orthogonal to the translations of the group $T$ or $T_h$,
while the translational periods of the groups $T$ and $T_h$ are not
related to those of $2_1$ (respectively $T_v$).

These five 2D generalized translational groups form the lattices
classified according to the four holohedries: the oblique (holohedry
$\C_{2h}$; arbitrary angle between the translational directions, with
different periods), the rectangular ($\D_{2h}$; orthogonal
translational directions with different periods), the square one
($\D_{4h}$; orthogonal translational directions and equal periods) and
the hexagonal one ($\D_{6h}$; the angle $2\pi /3$ between the
translational directions with the equal periods) \cite{KOPSKY}. If
together with the primitive rectangular translations $\vec{a}$ and
$\vec{b}$, the lattice contains the vector
$\frac{1}{2}(\vec{a}+\vec{b})$, it is called the centered rectangular,
to differ from the primitive ones (these generalized translational
groups are emphasized by prime in the text).

Depending on the type of the lattice, various orthogonal symmetries can
be involved. They combine into the point factors, being the axial point
groups, \cite{T.JANSEN}, leaving the $z$ axis invariant. Since the
crystallographic conditions on the order of the principal axis of
rotation must be imposed (analogously to the space groups, but in the
contrast to the line groups), the possible point factors are: $\C_n$,
$\C_{nv}$, $\C_{nh}$, $\D_n$, and $\D_{nh}$ for $n=1,\, 2,\, 3,\, 4$
and $6$; $\D_{nd}$ and ${\bf S}_{2n}$ for $n=1,\, 2$ and $3$. These are
also the possible isogonal point  groups, which are obtained by adding
the orthogonal part of the generalized translational generators to the
point factor (thus the point factor ${\bf P}$ is either the isogonal
point group, either its index-two subgroup).  The list of all
diperiodic groups (in the numerical, \cite{EW}, and international
notation), factorized in the described form ${\bf PZ}$, is given in the
table
\ref{PRVA}.

The factorization is utilized in the construction of the irrs.
Firstly, it immediately gives the generators of the diperiodic groups:
two generators for the generalized translational factor, ${\bf Z}$, and
at most three additional generators of the point factor ${\bf P}$.
Since the representation of a group is completely determined by the
matrices representing the generators, the irrs of the diperiodic groups
are in the next section tabulated by giving at most five matrices.
Further, the factorization straightforwardly gives optimal method for
the construction of irrs.  Namely, it enables to classify the groups
into the chains, so that each member of a chain is an index-2
subgroup of the next one. This is necessary in order to apply,
whenever it is possible, the simplest method of construction --- the
induction from the index-two subgroup, \cite{JABOON}.  The starting
group in each chain is either with known irrs (i. e. it belongs to some
other chain), or it is the direct or the semidirect product of its two
abelian subgroups, with the elaborated techniques of the construction of
irrs,  \cite{ALTMAN,JABOON}.  Hence, depending on the structure of the
diperiodic group, one of these three methods of constructing of their
irrs is applied. Some necessary details about these methods are briefly
sketched in the appendix.

{\renewcommand{\arraystretch}{1.2}{\footnotesize{
\begin{table}
\caption[hbt]{\label{PRVA}{\footnotesize{
{\bf The Factorization of the Diperiodic Groups.} 
For each diperiodic group $\D\g$, the holohedry ${\bf H}$, the isogonal
point group ${\bf I}$, the factorization ${\bf PT}$ and the
international symbol according to \cite{EW}, is given. The last column
refers to the table containing the irrs of the group.}}}
\begin{center}
\begin{tabular}{||c|c|c|c|c|c||c|c|c|c|c|c||} \hline \hline
$\D\g$&${\bf H}$&{\bf I}&{\bf P T}& Int. simb.&Table& 
$\D\g$&${\bf H}$&{\bf I}&{\bf P T}& Int. simb.&Table\\ \hline \hline
1&{\bf C}$_{2h}$  &${\bf C}_{1}$&$\T$& p1&\ref{1}&
41&${\bf D}_{2h}$&${\bf D}_{2h}$& ${\bf C}_{2v} \T_h$& p$\frac2a\frac{2_1}m\frac2m$&\ref{38}\\
\cline{1-1}\cline{3-7}\cline{9-12}
2&  &${\bf S}_{2}$&${\bf S}_2\T$& p$\bar{1}$&\ref{2}&
42& &${\bf D}_{2h}$& ${\bf D}_{1d}\T'_h$& p$\frac2n\frac2m\frac{2_1}a$&\ref{39}\\
\cline{1-1}\cline{3-7}\cline{9-12}
3& &${\bf C}_{2}$& {\bf C}$_2\T$& p211&\ref{2}&
43& &${\bf D}_{2h}$& ${\bf D}_12_1T_h$&  p$\frac2a\frac2b\frac{2_1}a$&\ref{43}\\
\cline{1-1}\cline{3-7}\cline{9-12}
4&  &${\bf C}_{1h}$& ${\bf C}_{1h}\T$& pm11&\ref{1}& 
44& &${\bf D}_{2h}$& ${\bf C}_{2h}\T'_v$& p$\frac2m\frac{2_1}b\frac{2_1}a$&\ref{18}\\
\cline{1-1}\cline{3-7}\cline{9-12}
5& &${\bf C}_{1h}$& $\T_h$& pb11&\ref{1}& 
45& &${\bf D}_{2h}$& ${\bf C}_2 2_1T_h$& p$\frac2a\frac{2_1}{b}\frac{2_1}{m}$&\ref{45}\\
\cline{1-1}\cline{3-7}\cline{9-12}
6& &${\bf C}_{2h}$& ${\bf C}_{2h}\T$& p$\frac{2}{m}$11&\ref{2}& 
46& &${\bf D}_{2h}$& ${\bf C}_{2v}\T'_h$& p$\frac2n\frac{2_1}m\frac{2_1}m$&\ref{39}\\
\cline{1-1}\cline{3-7}\cline{9-12}
7& &${\bf C}_{2h}$& ${\bf C}_2\T_h$& p$\frac2b$11&\ref{7}& 
47& &${\bf D}_{2h}$& ${\bf D}_{2h}\T'$&  c$\frac2m\frac2m\frac2m$&\ref{16}\\   
\cline{1-7}\cline{9-12}
8&{\bf D}$_{2h}$  &${\bf D}_{1}$& ${\bf D}_1\T$& p112&\ref{8}&
48& &${\bf D}_{2h}$& ${\bf C}_{2v}\T'_h$& c$\frac2a\frac2m\frac2m$&\ref{48}\\
\cline{1-1}\cline{3-12}
9& &${\bf D}_{1}$& ${\bf 2}_1$& p112$_1$&\ref{9}& 
49&${\bf D}_{4h}$ &${\bf C}_{4}$& ${\bf C}_4 \T$& p4&\ref{49}\\ 
\cline{1-1}\cline{3-7}\cline{9-12}
10& &${\bf D}_{1}$& ${\bf D}_1\T'$& c112&\ref{10}& 
50& &${\bf S}_{4}$& ${\bf S}_4\T$& p$\bar{4}$&\ref{49}\\
\cline{1-1}\cline{3-7}\cline{9-12}
11& &${\bf C}_{1v}$& ${\bf C}_{1v}\T$& p11m&\ref{8}&
51 & &${\bf C}_{4h}$& ${\bf C}_{4h}\T$& p4/m&\ref{49}\\ 
\cline{1-1}\cline{3-7}\cline{9-12}
12& &${\bf C}_{1v}$& $\T_v$& p11a&\ref{9}&
52& &${\bf C}_{4h}$& ${\bf C}_4\T'_h$& p4/n&\ref{52}\\ 
\cline{1-1}\cline{3-7}\cline{9-12}
13& &${\bf C}_{1v}$& ${\bf C}_{1v}\T'$& c11m&\ref{10}&
53& &${\bf D}_{4}$& ${\bf D}_4\T$& p422&\ref{53}\\
\cline{1-1}\cline{3-7}\cline{9-12}
14& &${\bf D}_{1d}$& ${\bf D}_{1d}\T$& p11$\frac2m$&\ref{14}&
54& &${\bf D}_{4}$& ${\bf C}_4{\bf 2}'_1$& p42$_1$2&\ref{54}\\
\cline{1-1}\cline{3-7}\cline{9-12}
15& &${\bf D}_{1d}$& ${\bf S}_2 {\bf 2}_1$& p11$\frac{2_1}m$&\ref{15}&
55& &${\bf C}_{4v}$& ${\bf C}_{4v}\T$& p4mm&\ref{53}\\ 
\cline{1-1}\cline{3-7}\cline{9-12}
16& &${\bf D}_{1d}$& ${\bf D}_{1d}\T'$& c11$\frac2m$&\ref{16}&
56& &${\bf C}_{4v}$& ${\bf C}_4\T'_v$& p4bm&\ref{54}\\
\cline{1-1}\cline{3-7}\cline{9-12}
17& &${\bf D}_{1d}$& ${\bf S}_2\T_v$& p11$\frac2a$&\ref{15}&
57& &${\bf D}_{2d}$& ${\bf D}_{2d}\T$& p$\bar{4}$2m&\ref{53}\\
\cline{1-1}\cline{3-7}\cline{9-12}
18& &${\bf D}_{1d}$& ${\bf S}_2 \T'_v$& p11$\frac{2_1}a$&\ref{18}&
58& &${\bf D}_{2d}$& ${\bf S}_4{\bf 2}'_1$& p$\bar{4}$2$_1$m&\ref{54}\\
\cline{1-1}\cline{3-7}\cline{9-12}
19& &${\bf D}_{2}$& ${\bf D}_2\T$& p222&\ref{14}&
59& &${\bf D}_{2d}$& ${\bf D}_{2d}\T$& p$\bar{4}$m2&\ref{53}\\
\cline{1-1}\cline{3-7}\cline{9-12}
20& &${\bf D}_{2}$& ${\bf C}_2 {\bf 2}_1$& p222$_1$&\ref{15}&
60& &${\bf D}_{2d}$& ${\bf S}_4\T'_v$& p$\bar4$b2&\ref{54}\\
\cline{1-1}\cline{3-7}\cline{9-12}
21& &${\bf D}_{2}$&${\bf C}_2 {\bf 2}'_1$& p22$_1$2$_1$&\ref{18}&
61& &${\bf D}_{4h}$& ${\bf D}_{4h}\T$& p$\frac4m\frac2m\frac2m$&\ref{53}\\
\cline{1-1}\cline{3-7}\cline{9-12}
22& &${\bf D}_{2}$& ${\bf D}_2\T'$& c222&\ref{16}&
62& &${\bf D}_{4h}$& ${\bf D}_{2d}\T'_h$& p$\frac4n\frac2b\frac2m$&\ref{62}\\
\cline{1-1}\cline{3-7}\cline{9-12}
23& &${\bf C}_{2v}$& ${\bf C}_{2v}\T$& p2mm&\ref{14}& 
63& &${\bf D}_{4h}$& ${\bf C}_{4h}\T'_v$& p$\frac4m\frac{2_1}b\frac2m$&\ref{54}\\
\cline{1-1}\cline{3-7}\cline{9-12}
24&  &${\bf D}_{1h}$& ${\bf D}_{1h}\T$& pmm2& \ref{8}& 
64& &${\bf D}_{4h}$& ${\bf D}_{2d}\T'_h$& p$\frac4n\frac{2_1}m\frac2m$&\ref{62}\\
\cline{1-1}\cline{3-12}
25& &${\bf D}_{1h}$& ${\bf C}_{1h}{\bf 2}_1$& pm2$_1$a&\ref{9}&
65&${\bf D}_{6h}$  &${\bf C}_{3}$&${\bf C}_3 \T$ & p3&\ref{65}\\ 
\cline{1-1}\cline{3-7}\cline{9-12}
26& &${\bf D}_{1h}$& ${\bf C}_{1v} {\bf 2}_1$& pbm2$_1$&\ref{9}&
66& &${\bf S}_{6}$& ${\bf S}_6\T$& p$\bar{3}$&\ref{66}\\
\cline{1-1}\cline{3-7}\cline{9-12}
27& &${\bf D}_{1h}$& ${\bf D}_1\T_v$& pbb2&\ref{9}&
67& &${\bf D}_{3}$& ${\bf D}_3\T$& p312&\ref{67}\\
\cline{1-1}\cline{3-7}\cline{9-12}
28& &${\bf C}_{2v}$& ${\bf C}_2 \T_v$& p2ma&\ref{15}&
68& &${\bf D}_{3}$& ${\bf D}_3\T$& p321&\ref{68}\\
\cline{1-1}\cline{3-7}\cline{9-12}
29& &${\bf D}_{1h}$& ${\bf D}_1\T_h$& pam2&\ref{29}&
69& &${\bf C}_{3v}$& ${\bf C}_{3v} \T$& p3m1&\ref{67}\\ 
\cline{1-1}\cline{3-7}\cline{9-12}
30& &${\bf D}_{1h}$& $2_{1}T_{h}$& pab2$_1$&\ref{29}&
70& &${\bf C}_{3v}$& ${\bf C}_{3v} \T$& p31m&\ref{68}\\ 
\cline{1-1}\cline{3-7}\cline{9-12}
31& &${\bf D}_{1h}$& ${\bf D}_1\T'_h$& pnb2&\ref{10}&
71& &${\bf D}_{3d}$& ${\bf D}_{3d}\T$& p$\bar{3}$1$\frac2m$&\ref{71}\\
\cline{1-1}\cline{3-7}\cline{9-12}
32& &${\bf D}_{1h}$& ${\bf C}_{1v}\T'_h$& p nm2$_1$&\ref{10}&
72& &${\bf D}_{3d}$& ${\bf D}_{3d}\T$& p$\bar{3}\frac{2}{m}$1&\ref{71}\\
\cline{1-1}\cline{3-7}\cline{9-12}
33& &${\bf C}_{2v}$& ${\bf C}_2\T'_v$ & p2ba&\ref{18}& 
73& &${\bf C}_{6}$& ${\bf C}_6 \T$& p6&\ref{66}\\
\cline{1-1}\cline{3-7}\cline{9-12}
34& &${\bf C}_{2v}$& ${\bf C}_{2v}\T'$& c2mm&\ref{16}& 
74& &${\bf C}_{3h}$& ${\bf C}_{3h}\T$& p$\bar{6}$&\ref{65}\\ 
\cline{1-1}\cline{3-7}\cline{9-12}
35& &${\bf D}_{1h}$& ${\bf D}_{1h}\T'$& cmm2&\ref{10}&
75& &${\bf C}_{6h}$& ${\bf C}_{6h}\T$& p6/m&\ref{66}\\
\cline{1-1}\cline{3-7}\cline{9-12}
36& &${\bf D}_{1h}$& ${\bf D}_1\T'_h$& cam2&\ref{36}&
76& &${\bf D}_{6}$& ${\bf D}_6\T$& p622&\ref{71}\\
\cline{1-1}\cline{3-7}\cline{9-12}
37& &${\bf D}_{2h}$& ${\bf D}_{2h}\T$& p$\frac2m\frac2m\frac2m$&\ref{14}& 
77& &${\bf C}_{6v}$& ${\bf C}_{6v}\T$& p6mm&\ref{71}\\ 
\cline{1-1}\cline{3-7}\cline{9-12}
38& &${\bf D}_{2h}$& ${\bf D}_2\T_h$& p$\frac2a\frac2m\frac2a$&\ref{38}&
78& &${\bf D}_{3h}$& ${\bf D}_{3h}\T$& p$\bar{6}$m2&\ref{67}\\ 
\cline{1-1}\cline{3-7}\cline{9-12}
39& &${\bf D}_{2h}$& ${\bf D}_2\T'_h$& p$\frac2n\frac2b\frac2a$&\ref{39}&
79& &${\bf D}_{3h}$& ${\bf D}_{3h}\T$& p$\bar{6}$2m&\ref{68}\\
\cline{1-1}\cline{3-7}\cline{9-12}
40& &${\bf D}_{2h}$& ${\bf C}_{2h}\T_v$& p$\frac2m\frac{2_1}m\frac2a$&\ref{15}&
80& &${\bf D}_{6h}$& ${\bf D}_{6h}\T$& p$\frac6m\frac2m\frac2m$&\ref{71}\\ 
\hline \hline
\end{tabular}
\end{center}
\end{table}}}}

\section{The irreducible representations (irrs)}\label{MAIN}
In this section, the irrs of the diperiodic groups are tabulated.  The
most of the tables present the irrs of several diperiodic groups. If
the group {\bf Dg} is generated by the set $\{g_1,g_2,\dots\}$, this is
denoted in the caption as ${\bf Dg}=\gr\{g_1,g_2,\dots\}$.  The symbols
of the generators are: $C_n$ is the rotation for $\frac{2\pi}{n}$
around $z$-axis, $\gs_h$ is the horizontal mirror plane, while
$\gs_x$, $\gs_y$ and $\gs$ are the vertical mirror planes containing
the $x$-axis, $y$-axis and the axis $x=y$, respectively; rotations for
$\pi$ around $x$-axis, $y$-axis and the line $x=y$ are denoted by
$U_x$, $U_y$ and $U$. The Koster-Seitz notation is used for the
generators of the generalized translations: $(A|x y)$ is the orthogonal
transformation $A$ followed by the translations for $x$ and $y$ along
the corresponding directions.
 
To find the representations of the group {\bf Dg}, only the matrices
corresponding to these generators are to be taken.  Each raw of a table
gives one or more irrs of the groups enumerated in the caption.  In the
first column, the symbol of the representation is indicated, and its
dimension follows in the column 2.  The matrices of the generators are
listed in the remaining columns.  Some additional explanations, given
in the captions of the tables, are necessary to describe the specific
notion and the range of the quantum numbers.

The general label of the representation, ${_{\bf k}^{v,t}D^\pm_m}$,
emphasizes the symmetry based quantum numbers of the corresponding
states.  The left subscript ${\bf k}$ is the wave vector, taking the
values from the irreducible (basic) domain of the Brillouin zone
(shaded part in the figures). The Brillouin zones are chosen as the
oblique, rectangular, square and hexagonal. There are eight types of
the irreducible domains, \cite{ALTMAN2}, drawn in the figures
\ref{Oid}-\ref{HEXid}, and each table refers to one of these domains,
specified in the caption.  It is either the whole Brillouin zone, or
the part of it ($\frac{1}{2}$, $\frac{1}{3}$, $\frac{1}{4}$,
$\frac{1}{6}$, $\frac{1}{8}$, $\frac{1}{12}$).  Only the bold lines and
the filled points at the boundary of the basic domain belong to the
domain (when necessary, the white circles specify the boundary points
excluded from the domain).  If the boundary point or line is special,
i. e. with the representations differing from those in the adjacent
points, it is additionally labeled at the corresponding figure. The
basic symbol, $D$, specifies the type of the position in the zone: $G$
stands for the general interior point of the domain, ${\bf
k}=(k_1,k_2)$, while the other letters denote the special points
($\Gamma$, $X$, $Y$, $M$, $Q$ and $Q'$) and the special lines
($\Delta$, $\Lambda$, $\Sigma$, $\Sigma'$, $\Upsilon$ and $\Phi$). Note
that the translational periods are used as the unit lengths along the
corresponding directions.

The components $k_1$ and $k_2$ of the quasi momentum vector ${\bf k}$
are conjugated to the translational directions of the 2D generalized
translational group. For rectangular and square diperiodic groups,
$k_1$ and $k_2$ are the Cartesian coordinates $k_x$ and $k_y$,
respectively.  The right subscript $m$ is the quasi angular momentum;
it takes on the integer values specified in the table captions. The
quantum numbers $v$ and $t$ take on the values 0 and 1. The first one
is the parity of the mirror symmetry in the vertical planes, or of
the rotation for $\pi$
around the horizontal axes. The second one,  $t$,
refers to the rectangular centered groups only, being related to the
element $(I|\frac{1}{2}\,\frac{1}{2})$.  The $\pm$ signs are reserved
for the symmetry of the horizontal plane $\sigma_h$, or of the glide
plane $(\gs_h|\frac{1}{2} \frac{1}{2})$.

The method of construction of the irrs is indicated in the caption.
As a rule, the irrs of the translational subgroup are found first, then
the rotations around $z$-axis are included (the translations being
an abelian invariant subgroup). This group and its irrs are the starting
point for the chain of successive inductions (from the index-two
subgroup) procedures, until the whole group being incorporated. 

Some abbreviations, necessary to make the tables transparent, are listed
separately for each type of the groups. Throughout the text, the $n$
dimensional identity matrix is denoted by $I_n$, while $A_n$ stands for
the offdiagonal matrix $A_n={\rm offdiag}\, (\underbrace{1,\ldots
,1}_n)$.

\subsection{The irrs of the oblique groups}
The diperiodic groups with the primitive translations making an
arbitrary angle are ${\bf Dg}1$ - ${\bf Dg}7$.  Their irrs are listed
in the tables \ref{1}-\ref{7}.  Since the groups with different angles
between the translational periods are isomorphic, their representations
are same; therefore, although at the figure \ref{Oid}, the rectangular
irreducible domains are depicted, suitable for the construction of
the irrs of other groups, the figures refer equally well to the most
general case. 
The diperiodic groups ${\bf Dg}1$, ${\bf Dg}4$ and ${\bf Dg}5$ are the
direct products and their irrs are obtained as the
products of the relevant subgroup irrs. 

{\footnotesize
\begin{figure}[hbt]\begin{center}
\unitlength=0.80mm
\linethickness{0.4pt}
\begin{picture}(115.00,70.00)

\put(25.00,10.00){\makebox(0,0)[cc]{$(a)$}}
\linethickness{1pt}
\put(40.00,60.00){\line(-1,0){29.00}}
\put(40.00,60.00){\line(0,-1){39.00}}
\linethickness{0.1pt}
\put(25.00,40.00){\vector(1,0){28.00}}
\put(25.00,40.00){\vector(0,1){30.00}}
\put(21.00,65.00){\makebox(0,0)[cc]{$k_2$}}
\put(49.00,36.00){\makebox(0,0)[cc]{$k_1$}}

\put(10.00,20.00){\line(1,0){29.00}}
\put(10.00,20.00){\line(0,1){39.00}}
\put(10.00,22.00){\line(1,0){30.00}}
\put(10.00,24.00){\line(1,0){30.00}}
\put(10.00,26.00){\line(1,0){30.00}}
\put(10.00,28.00){\line(1,0){30.00}}
\put(10.00,30.00){\line(1,0){30.00}}
\put(10.00,32.00){\line(1,0){30.00}}
\put(10.00,34.00){\line(1,0){30.00}}
\put(10.00,36.00){\line(1,0){30.00}}
\put(10.00,38.00){\line(1,0){30.00}}
\put(10.00,40.00){\line(1,0){30.00}}
\put(10.00,42.00){\line(1,0){30.00}}
\put(10.00,44.00){\line(1,0){30.00}}
\put(10.00,46.00){\line(1,0){30.00}}
\put(10.00,48.00){\line(1,0){30.00}}
\put(10.00,50.00){\line(1,0){20.00}}
\put(35.00,50.00){\line(1,0){5.00}}
\put(10.00,52.00){\line(1,0){20.00}}
\put(35.00,52.00){\line(1,0){5.00}}
\put(10.00,54.00){\line(1,0){20.00}}
\put(35.00,54.00){\line(1,0){5.00}}
\put(10.00,56.00){\line(1,0){30.00}}
\put(10.00,58.00){\line(1,0){30.00}}
\put(10.00,60.00){\circle{2.00}}
\put(40.00,20.00){\circle{2.00}}
\put(32.00,52.00){\makebox(0,0)[cc]{$G$}}
\put(95.00,10.00){\makebox(0,0)[cc]{$(b)$}}
\linethickness{1pt}
\put(110.00,60.00){\line(-1,0){29.00}}
\put(110.00,40.00){\line(-1,0){29.00}}
\put(110.00,60.00){\line(0,-1){20.00}}
\linethickness{0.1pt}
\put(95.00,40.00){\vector(1,0){28.00}}
\put(95.00,40.00){\vector(0,1){30.00}}
\put(91.00,67.00){\makebox(0,0)[cc]{$k_2$}}
\put(119.00,38.00){\makebox(0,0)[cc]{$k_1$}}

\put(80.00,20.00){\line(1,0){30.00}}
\put(110.00,20.00){\line(0,1){20.00}}
\put(80.00,20.00){\line(0,1){19.00}}
\put(80.00,41.00){\line(0,1){18.00}}
\put(80.00,42.00){\line(1,0){30.00}}
\put(80.00,44.00){\line(1,0){30.00}}
\put(80.00,46.00){\line(1,0){30.00}}
\put(80.00,48.00){\line(1,0){30.00}}
\put(80.00,50.00){\line(1,0){20.00}}
\put(105.00,50.00){\line(1,0){5.00}}
\put(80.00,52.00){\line(1,0){20.00}}
\put(105.00,52.00){\line(1,0){5.00}}
\put(80.00,54.00){\line(1,0){20.00}}
\put(105.00,54.00){\line(1,0){5.00}}

\put(80.00,56.00){\line(1,0){30.00}}
\put(80.00,58.00){\line(1,0){30.00}}
\put(80.00,60.00){\circle{2.00}}
\put(80.00,40.00){\circle{2.00}}
\put(110.00,60.00){\circle*{2.00}}
\put(95.00,60.00){\circle*{2.00}}
\put(95.00,40.00){\circle*{2.00}}
\put(110.00,40.00){\circle*{2.00}}
\put(102.00,52.00){\makebox(0,0)[cc]{$G$}}

\put(96.00,64.00){\makebox(0,0)[lc]{$Y$}}
\put(110.00,64.00){\makebox(0,0)[cc]{$M$}}
\put(113.00,41.00){\makebox(0,0)[cb]{$X$}}
\put(95.00,36.00){\makebox(0,0)[cc]{$\Gamma$}}
\end{picture}
\caption[]{\label{Oid}{\footnotesize 
The irreducible domains of the {\Brillouin} zone for the oblique
diperiodic groups.  The coordinates of the special points (filled
circles) are: $\gG =(0,0)$, $X=(\pi ,0)$, $Y=(0,\pi )$ and $M=(\pi ,\pi
)$.  The coordinate lines $k_1$ and $k_2$ are allowed not to be orthogonal.}}
\end{center}\end{figure}
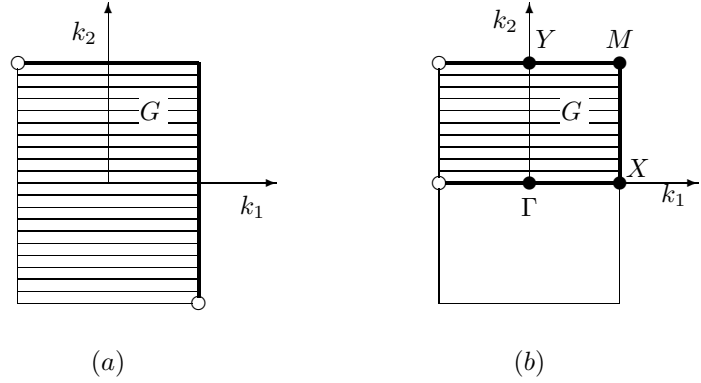}

{\footnotesize{
\begin{table}[hbt]
\caption[]{\label{1}{\footnotesize{The irrs of the oblique 2D translational 
group $\D\g 1=\T =\gr\{\tx ,\ty\}$ and the oblique generalized 2D
translational groups $\D\g 4 =\C_{1h}\T =\gr\{\gs_h, \tx, \ty\}$ and
$\D\g 5 =\T_h=\gr\{\tx, \zhy\}$.  The {\Brillouin} zone, being the
irreducible domain is shown in the Fig. \ref{Oid}$\, (a)$.}}}
\begin{center}
\begin{tabular}{||c|c||c|c|c|c||} \hline \hline 
Irr&D&$\gs_h$&$(I|1\,0)$&$(I|0\,1)$&$(\sigma_h|0\,\frac{1}{2})$\\ \hline\hline
$_{\bf k}G^{\pm}$&1&$\pm$1&$e^{\imath k_1}$&$e^{\imath k_2}$
&$\pm e^{\imath\frac{k_2}{2}}$\\ \hline \hline 
\end{tabular}\end{center}\end{table}}}
{\footnotesize{
\begin{table}[hbt]
\caption[]{\label{2}{\footnotesize{The irrs of the oblique diperiodic groups  
$\D\g 2={\bf S}_2\T =\gr\{C_2\gs_h, \tx, \ty\}$, 
$\D\g 3={\bf C}_2\T =\gr\{C_2, \tx, \ty\}$ and 
$\D\g 6={\bf C}_{2h}\T =\gr\{C_2, \gs_h, \tx, \ty\}$ 
induced by the elements $C_2\sigma_h$ or $C_2$ from the irrs of the groups 
${\bf Dg}1$ and ${\bf Dg}4$ (Tab. \ref{1}). 
The quasi angular momentum $m$ takes on the values $0$ and $1$. 
The irreducible domain is presented in the Fig. \ref{Oid}$\, (b)$.}}}
\begin{center}
\begin{tabular}{||c|c||c|c|c|c||} \hline \hline 
Irr&D&$C_2$ or $C_2\sigma_h$&$\gs_h$&$(I|1\,0)$&$(I|0\,1)$\\ \hline\hline
$\gG_m^{\pm}$&1&(-1)$^m$&$\pm 1$&1&1\\ \hline 
$\X_m^{\pm}$&1&(-1)$^m$&$\pm 1$&-1&1\\ \hline 
$\Y_m^{\pm}$&1&(-1)$^m$&$\pm 1$&1&-1\\ \hline 
$\M_m^{\pm}$&1&(-1)$^m$&$\pm 1$&-1&-1  \\ \hline 
$_{\bf k}G^{\pm}$&2&$A_2$&$\pm I_2$&
$\pmatrix{e^{\imath k_1}&0\\0&e^{-\imath k_1}\\}$&
$\pmatrix{e^{\imath k_2}&0\\0&e^{-\imath k_2}\\}$\\ \hline \hline 
\end{tabular}\end{center}\end{table}}}
{\footnotesize{
\begin{table}[hbt]
\caption[]{\label{7}{\footnotesize{The irrs of the oblique diperiodic group
$\D\g 7={\bf C}_2\T_h=\gr\{C_2, \tx, \zhy\}$ induced by the element 
$(\gs_h|0\,\frac{1}{2})$ from the irrs of
the group ${\bf Dg}3$ (Tab. \ref{2}). The quasi angular momentum $m$ 
takes on the values $0$ and $1$.
The irreducible domain is presented in the Fig. \ref{Oid}$\, (b)$.}}}
\begin{center}
\begin{tabular}{||c|c||c|c|c||} \hline \hline 
Irr&D&$C_2$&$(\gs_h|0\,\frac{1}{2})$&$(I|1\,0)$\\ \hline\hline
$\gG_m^\pm$&1&(-1)$^m$&$\pm 1$&1\\ \hline 
$\X_m^\pm$&1&(-1)$^m$&$\pm$1&-1\\ \hline 
$Y$&2&$A_2$&$\pmatrix{\imath&0\\0&-\imath\\}$&$I_2$\\ \hline
$M$&2&$A_2$&$\pmatrix{-\imath&0\\0&\imath\\}$&$-I_2$\\ \hline
$_{\bf k}G^\pm$&2&$A_2$&
$\pm \pmatrix{e^{\imath\frac{k_2}{2}}&0\\0&e^{-\imath\frac{k_2}{2}}\\}$ 
&$\pmatrix{e^{\imath k_1}&0\\0&e^{-\imath k_1}\\}$\\ \hline \hline 
\end{tabular}\end{center}\end{table}}}

\clearpage
\subsection{The irrs of the rectangular groups}
The groups ${\bf Dg}8$ - ${\bf Dg}48$ are rectangular. Among these 41
groups there are 23 with the primitive and 18 with the
centered lattice.  The irreducible domains are given in the Fig.
\ref{PRIMid}.  The irrs are induced from the irrs of the oblique
groups with the rectangular translational directions.
Firstly, the irrs of the rectangular primitive groups are
obtained and afterwards the induction by the nonsymorphic generators
$(I|\frac{1}{2}\,\frac{1}{2})$, $(\sigma_h|\frac{1}{2}\,\frac{1}{2})$,
$(\sigma |\frac{1}{2}\,\frac{1}{2})$ and $(U|\frac{1}{2}\,\frac{1}{2})$
gives the irrs of the rectangular centered groups.

{\footnotesize
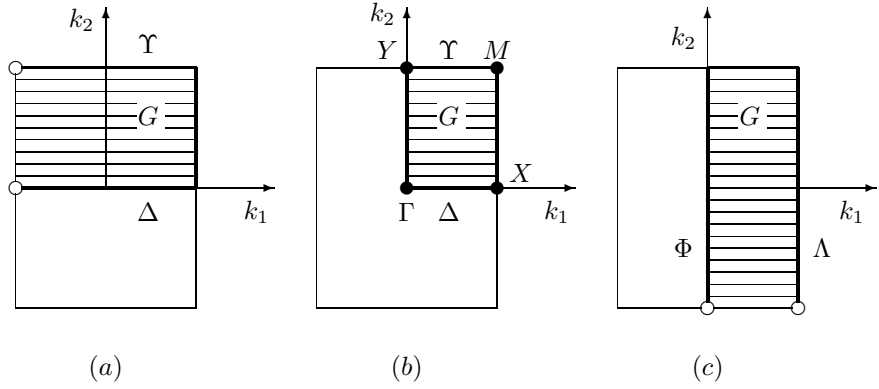
\begin{figure}[hbt]\begin{center}
\unitlength=0.80mm
\linethickness{0.4pt}
\begin{picture}(153.00,80)

\put(25.00,10.00){\makebox(0,0)[cc]{$(a)$}}
\linethickness{1pt}
\put(40.00,60.00){\line(-1,0){29.00}}
\put(40.00,40.00){\line(-1,0){29.00}}
\put(40.00,60.00){\line(0,-1){20.00}}
\linethickness{0.1pt}
\put(25.00,40.00){\vector(1,0){28.00}}
\put(25.00,40.00){\vector(0,1){30.00}}
\put(21.00,68.00){\makebox(0,0)[cc]{$k_2$}}
\put(50.00,36.00){\makebox(0,0)[cc]{$k_1$}}

\put(10.00,20.00){\line(1,0){30.00}}
\put(40.00,20.00){\line(0,1){20.00}}
\put(10.00,20.00){\line(0,1){19.00}}
\put(10.00,41.00){\line(0,1){18.00}}
\put(10.00,42.00){\line(1,0){30.00}}
\put(10.00,44.00){\line(1,0){30.00}}
\put(10.00,46.00){\line(1,0){30.00}}
\put(10.00,48.00){\line(1,0){30.00}}

\put(10.00,50.00){\line(1,0){20.00}}
\put(35.00,50.00){\line(1,0){5.00}}
\put(10.00,52.00){\line(1,0){20.00}}
\put(35.00,52.00){\line(1,0){5.00}}
\put(10.00,54.00){\line(1,0){20.00}}
\put(35.00,54.00){\line(1,0){5.00}}

\put(10.00,56.00){\line(1,0){30.00}}
\put(10.00,58.00){\line(1,0){30.00}}
\put(10.00,60.00){\circle{2.00}}
\put(10.00,40.00){\circle{2.00}}
\put(32.00,36.00){\makebox(0,0)[cc]{$\Delta$}}
\put(32.00,64.00){\makebox(0,0)[cc]{$\Upsilon$}}
\put(32.00,52.00){\makebox(0,0)[cc]{$G$}}
\put(75.00,10.00){\makebox(0,0)[cc]{$(b)$}}
\linethickness{1pt}
\put(90.00,60.00){\line(-1,0){15.00}}
\put(90.00,40.00){\line(-1,0){15.00}}
\put(90.00,60.00){\line(0,-1){20.00}}
\put(75.00,60.00){\line(0,-1){20.00}}
\linethickness{0.1pt}
\put(75.00,40.00){\vector(1,0){28.00}}
\put(75.00,40.00){\vector(0,1){30.00}}
\put(71.00,69.00){\makebox(0,0)[cc]{$k_2$}}
\put(100.00,36.00){\makebox(0,0)[cc]{$k_1$}}

\put(60.00,60.00){\line(1,0){15.00}}
\put(60.00,20.00){\line(1,0){30.00}}
\put(90.00,20.00){\line(0,1){20.00}}
\put(60.00,20.00){\line(0,1){40.00}}
\put(75.00,42.00){\line(1,0){15.00}}
\put(75.00,44.00){\line(1,0){15.00}}
\put(75.00,46.00){\line(1,0){15.00}}
\put(75.00,48.00){\line(1,0){15.00}}
\put(75.00,50.00){\line(1,0){5.00}}
\put(85.00,50.00){\line(1,0){5.00}}
\put(75.00,52.00){\line(1,0){5.00}}
\put(85.00,52.00){\line(1,0){5.00}}
\put(75.00,54.00){\line(1,0){5.00}}
\put(85.00,54.00){\line(1,0){5.00}}
\put(75.00,56.00){\line(1,0){15.00}}
\put(75.00,58.00){\line(1,0){15.00}}

\put(90.00,60.00){\circle*{2.00}}
\put(75.00,60.00){\circle*{2.00}}
\put(75.00,40.00){\circle*{2.00}}
\put(90.00,40.00){\circle*{2.00}}
\put(70.00,63.00){\makebox(0,0)[lc]{$Y$}}
\put(90.00,63.00){\makebox(0,0)[cc]{$M$}}
\put(94.00,41.00){\makebox(0,0)[cb]{$X$}}
\put(75.00,36.00){\makebox(0,0)[cc]{$\Gamma$}}
\put(82.00,36.00){\makebox(0,0)[cc]{$\Delta$}}
\put(82.00,63.00){\makebox(0,0)[cc]{$\Upsilon$}}
\put(82.00,52.00){\makebox(0,0)[cc]{$G$}}
\put(125.00,10.00){\makebox(0,0)[cc]{$(c)$}}
\linethickness{1pt}
\put(140.00,60.00){\line(-1,0){15.00}}
\put(140.00,60.00){\line(0,-1){39.00}}
\put(125.00,60.00){\line(0,-1){39.00}}
\linethickness{0.1pt}
\put(125.00,40.00){\vector(1,0){28.00}}
\put(125.00,40.00){\vector(0,1){30.00}}
\put(121.00,65.00){\makebox(0,0)[cc]{$k_2$}}
\put(149.00,36.00){\makebox(0,0)[cc]{$k_1$}}

\put(110.00,60.00){\line(1,0){15.00}}
\put(110.00,20.00){\line(1,0){14.00}}
\put(126.00,20.00){\line(1,0){13.00}}
\put(110.00,20.00){\line(0,1){40.00}}
\put(125.00,22.00){\line(1,0){15.00}}
\put(125.00,24.00){\line(1,0){15.00}}
\put(125.00,26.00){\line(1,0){15.00}}
\put(125.00,28.00){\line(1,0){15.00}}
\put(125.00,30.00){\line(1,0){15.00}}
\put(125.00,32.00){\line(1,0){15.00}}
\put(125.00,34.00){\line(1,0){15.00}}
\put(125.00,36.00){\line(1,0){15.00}}
\put(125.00,38.00){\line(1,0){15.00}}
\put(125.00,40.00){\line(1,0){15.00}}
\put(125.00,42.00){\line(1,0){15.00}}
\put(125.00,44.00){\line(1,0){15.00}}
\put(125.00,46.00){\line(1,0){15.00}}
\put(125.00,48.00){\line(1,0){15.00}}
\put(125.00,50.00){\line(1,0){5.00}}
\put(135.00,50.00){\line(1,0){5.00}}
\put(125.00,52.00){\line(1,0){5.00}}
\put(135.00,52.00){\line(1,0){5.00}}
\put(125.00,54.00){\line(1,0){5.00}}
\put(135.00,54.00){\line(1,0){5.00}}
\put(125.00,56.00){\line(1,0){15.00}}
\put(125.00,58.00){\line(1,0){15.00}}
\put(125.00,20.00){\circle{2.00}}
\put(140.00,20.00){\circle{2.00}}
\put(132.00,52.00){\makebox(0,0)[cc]{$G$}}
\put(121.00,30.00){\makebox(0,0)[cc]{$\Phi$}}
\put(144.00,30.00){\makebox(0,0)[cc]{$\Lambda$}}
\linethickness{1pt}
\linethickness{0.1pt}
\end{picture}
\caption[]{\label{PRIMid}{\footnotesize 
The irreducible domains of the {\Brillouin} zone for the rectangular
diperiodic groups. The different special points (filled circles) are
$\gG =(0,0)$, $X=(\pi ,0)$, $Y=(0,\pi )$ and $M=(\pi ,\pi )$, while the
special lines are $\Delta =(k,0)$, $\Upsilon =(k,\pi )$, $\Phi =(0,k)$,
$\Upsilon =(\pi ,k)$, $\Sigma =(k,k)$ and $\Sigma '=(-k,k)$.}}
\end{center}\end{figure}}

{\footnotesize{
\begin{table}[hbt]
\caption[]{\label{8}{\footnotesize{The irrs of the rectangular
primitive diperiodic groups $\D\g 8={\bf D}_1\T =\gr\{U_x, \tx, \ty\}$,
$\D\g 11={\bf C}_{1v}\T =\gr\{\gs_x, \tx, \ty\}$ and 
$\D\g 24={\bf D}_{1h}\T =\gr\{U_x, \gs_h, \tx, \ty\}$ induced by
the elements $U_x$ and $\sigma_x$ from the irrs of the
groups ${\bf Dg}1$ and ${\bf Dg}4$ (Tab. \ref{1}).
The irreducible domain is presented in the
Fig. \ref{PRIMid}$\, (a)$.}}}
\begin{center}
\begin{tabular}{||c|c||c|c|c|c||} \hline \hline 
Irr&D&$\gs_x\ {\rm or}\ U_x$&$\gs_h$&$(I|1\,0)$&$(I|0\,1)$\\ \hline\hline
$_{k}^v\gD^\pm$&1&(-1)$^v$&$\pm 1$&$e^{\imath k}$&1\\ \hline 
$_{k}^v\Upsilon^\pm$&1&(-1)$^v$&$\pm 1$&$e^{\imath k}$&-1\\ \hline 
$_{\bf k}G^\pm$&2&$A_2$&$\pm I_2$&
$e^{\imath k_x}I_2$&$\pmatrix{e^{\imath k_y}&0\\0&e^{-\imath k_y}\\}$\\
\hline \hline 
\end{tabular}\end{center}\end{table}}}
{\footnotesize{
\begin{table}[hbt]
\caption[]{\label{14}{\footnotesize{
The irrs of the rectangular primitive diperiodic groups 
$\D\g 14={\bf D}_{1d}\T =\gr\{C_2\gs_h, \gs_x, \tx, \ty\}$, 
$\D\g 19={\bf D}_2\T =\gr\{C_2, U_x, \tx, \ty\}$,
$\D\g 23={\bf C}_{2v}\T =\gr\{C_2, \gs_x, \tx, \ty\}$ and 
$\D\g 37={\bf D}_{2h}\T =\gr\{C_2, U_x, \gs_h, \tx, \ty\}$ 
induced by the elements $U_x$ and $\sigma_x$ 
from the irrs of the groups ${\bf Dg}2$, ${\bf Dg}3$ and  
${\bf Dg}6$ (Tab. \ref{2}). The irreducible domain is presented in the 
Fig. \ref{PRIMid}$\, (b)$. The quasi angular momentum $m$ takes on the values
$0$ and $1$. Here, $K_2=\diag{(e^{\imath k_x},e^{-\imath k_x})}$.}}}
\begin{center}
\begin{tabular}{||c|c||c|c|c|c|c||} \hline \hline 
Irr&D&$C_2$ or $C_2\sigma_h$&$\gs_x\ {\rm or}\ U_x$&$\gs_h$&
$(I|1\,0)$&$(I|0\,1)$ \\ \hline\hline
$^v\gG^\pm_m$&1&(-1)$^m$&$(-1)^v$&$\pm 1$&1&1\\ \hline 
$^v\X^\pm_m$&1&(-1)$^m$&$(-1)^v$&$\pm 1$&-1&1\\ \hline 
$^v\Y^\pm_m$&1&(-1)$^m$&$(-1)^v$&$\pm 1$&1&-1\\ \hline 
$^v\M^\pm_m$&1&(-1)$^m$&$(-1)^v$&$\pm 1$&-1&-1  \\ \hline 
$_{k}^v\gD^\pm$&2&$A_2$&$(-1)^vI_2$&$\pm I_2$& 
$\pmatrix{e^{\imath k}&0\\0&e^{-\imath k}\\}$&$I_2$\\ \hline 
$_{k}^v\Upsilon^\pm$&2&$A_2$&$(-1)^vI_2$&$\pm I_2$& 
$\pmatrix{e^{\imath k}&0\\0&e^{-\imath k}\\}$&$-I_2$\\ \hline 
$_{k}^v\Phi^\pm$&2&$A_2$&$(-1)^vA_2$&$\pm I_2$& 
$I_2$&$\pmatrix{e^{\imath k}&0\\0&e^{-\imath k}\\}$\\ \hline 
$_{k}^v\Lambda^\pm$&2&$A_2$&$(-1)^vA_2$&$\pm I_2$& 
$-I_2$&$\pmatrix{e^{\imath k}&0\\0&e^{-\imath k}\\}$\\ \hline 
$_{\bf k}G^\pm$&4&$\pmatrix{A_2&0\\0&A_2\\}$&
$\pmatrix{0&I_2\\I_2&0\\}$ & $\pm I_4$&
$\pmatrix{K_2&0\\0&K_2\\}$&$\pmatrix{e^{\imath k_y}&0&0\\
0&e^{-\imath k_y}I_2&0\\ 0&0&e^{\imath k_y}\\}$\\ \hline \hline 
\end{tabular}\end{center}\end{table}}}
{\footnotesize{
\begin{table}[hbt]
\caption[]{\label{9}{\footnotesize{The irrs of the rectangular primitive 
diperiodic groups 
$\D\g 9 = {\bf 2}_1=\gr\{\tx, \zuy\}$, 
$\D\g 12 =\T_v=\gr\{\tx, \zvy\}$, 
$\D\g 25 ={\bf C}_{1h}{\bf 2}_1=\gr\{\gs_h, \tx, \zuy\}$,
$\D\g 26 ={\bf C}_{1v}{\bf 2}_1=\gr\{\gs_y, \tx, \zuy\}$ and 
$\D\g 27= {\bf D}_1\T_v=\gr\{U_y, \tx, \zvy\}$.
The irrs of the first three groups are 
induced by the elements
$(U_y|0\,\frac{1}{2})$ and $(\sigma_y|0\,\frac{1}{2})$ from
the groups ${\bf Dg}1$ and ${\bf Dg}4$ (Tab. \ref{1}), while 
the irrs of $\D\g 26$ and $\D\g 27$ are 
induced by the elements 
$\sigma_y$ and $U_y$ from the
irrs of the groups ${\bf Dg}9$ and ${\bf Dg}12$.
The irreducible domain is presented in the Fig. \ref{PRIMid}$\, (c)$.}}}
\begin{center}
{\tabcolsep1truemm  
\begin{tabular}{||c|c||c|c|c|c||} \hline\hline
Irr&D&$\gs_h$&$\gs_y\ {\rm or}\ U_y$&$(\gs_y|0\,\frac{1}{2})\ {\rm or}\ (U_y|0\,\frac{1}{2})$&
$(I|1\, 0)$\\ \hline \hline
$_{k}^v\Phi^\pm$&1&$\pm 1$&$(-1)^{v_1}$&$(-1)^ve^{\imath\frac{k}{2}}$&1\\ \hline
$_{k}^v\gL^\pm$&1&$\pm 1$&$(-1)^{v_1}$&
$(-1)^ve^{\imath\frac{k}{2}}$&-1\\ \hline
$_{\bf k}G^\pm$&2&$\pm I_2$
&$(-1)^v\pmatrix{0&e^{\imath\frac{k_y}{2}}\cr 
e^{-\imath\frac{k_y}{2}}&0\cr}$&$\pmatrix{0&e^{\imath k_y}\\1&0\\}$&
$\pmatrix{e^{\imath k_x}&0\\0&e^{-\imath k_x}\\}$\\ \hline\hline
\end{tabular}}\end{center}\end{table}}}
{\footnotesize{
\begin{table}[hbt]
\caption[]{\label{15}{\footnotesize{The irrs of the rectangular
primitive diperiodic groups  
$\D\g 15={\bf S}_2{\bf 2}_1=\gr\{C_2\gs_h, \tx, \zuy\}$, 
$\D\g 17={\bf S}_2\T_v=\gr\{C_2\gs_h, \tx, \zvy\}$,
$\D\g 20={\bf C}_2{\bf 2}_1=\gr\{C_2, \tx, \zuy\}$, 
$\D\g 28={\bf C}_2\T_v=\gr\{C_2, \tx, \zvy\}$ and 
$\D\g 40={\bf C}_{2h}\T_v=\gr\{C_2, \gs_h, \tx, \zvy\}$ 
induced by the element $C_2$ or $\sigma_hC_2$ 
from the irrs of the groups ${\bf Dg}9$, ${\bf Dg}12$ and ${\bf Dg}25$
(Tab. \ref{9}). The irreducible domain is presented in the  
Fig. \ref{PRIMid}$\, (b)$. The quasi angular momentum $m$ takes on the values
$0$ and $1$. Here, $L_2=\pmatrix{0&e^{\imath k_y}\cr 1&0\cr}$.}}}
\begin{center}
{\tabcolsep1truemm  
\begin{tabular}{||c|c||c|c|c|c||} \hline\hline
Irr&D&$C_2$ or $C_2\sigma_h$&$\gs_h$&
$(\gs_y|0\,\frac{1}{2})\ {\rm or}\ (U_y|0\,\frac{1}{2})$&
$(I|1\, 0)$\\ \hline \hline
$^v\gG_m^\pm$&1&$(-1)^m$&$\pm 1$&$(-1)^v$&$1$ \\ \hline
$^vX_m^\pm$&1&$(-1)^m$&$\pm 1$&$(-1)^v$&$-1$ \\ \hline
$Y^\pm$&2&$A_2$&$\pm I_2$&$\pmatrix{\imath&0\cr 0&-\imath\cr}$&$I_2$\\ \hline
$M^\pm$&2&$A_2$&$\pm I_2$&$\pmatrix{\imath&0\cr 0&-\imath\cr}$&$-I_2$\\ \hline
$^v_k\Phi^\pm$&2&$A_2$&$\pm I_2$&$(-1)^v
\pmatrix{e^{\imath\frac{k}{2}}&0\cr 0&e^{-\imath\frac{k}{2}}\cr}$&
$I_2$\\ \hline
$^v_k\Lambda^\pm$&2&$A_2$&$\pm I_2$&$(-1)^v
\pmatrix{e^{\imath\frac{k}{2}}&0\cr 0&e^{-\imath\frac{k}{2}}\cr}$&
$-I_2$\\ \hline
$_k\Delta_m^\pm$&2&$(-1)^mA_2$&$\pm I_2$&$A_2$&$
\pmatrix{e^{\imath k}&0\cr 0&e^{-\imath k}\cr}$\\ \hline
$_k\Upsilon_m^\pm$&2&$(-1)^mA_2$&$\pm I_2$&$\pmatrix{0&-1\cr 1&0\cr}$&$
\pmatrix{e^{\imath k}&0\cr 0&e^{-\imath k}\cr}$\\ \hline
$_{\bf k}G^\pm$&4&$\pmatrix{0&I_2\\I_2&0\\}$&$\pm I_4$& 
$\pmatrix{L_2&0\\0&L_2^{-1}\\}$&$\pmatrix{e^{\imath k_x}&0&0\cr 
0&e^{-\imath k_x}I_2&0\cr 0&0&e^{\imath k_x}\cr}$\\ \hline\hline
\end{tabular}}\end{center}\end{table}}}
{\footnotesize{
\begin{table}[hbt]
\caption[]{\label{29}{\footnotesize{
The irrs of the rectangular primitive diperiodic groups
$\D\g 29={\bf D}_1\T_h=\gr\{U_x, \tx, \zhy\}$ and
$\D\g 30=2_1T_h=\gr\{\zux, \zhy\}$
induced by the elements $(U_x|\frac{1}{2}\, 0)$ and $U_x$ from the
irrs of the group ${\bf Dg}5$ (Tab. \ref{1}). 
The irreducible domain is presented in the Fig. \ref{PRIMid}$\, (a)$.}}}
\begin{center}
{\tabcolsep1truemm  
\begin{tabular}{||c|c||c|c|c|c||} \hline\hline
Irr&D&$U_x$&$(U_x|\frac{1}{2}\, 0)$&$(\gs_h|0\,\frac{1}{2})$&$(I|1\, 0)$
\\ \hline \hline
$_{k}^v\gD^\pm$&1&$(-1)^v$&$(-1)^ve^{\imath\frac{k}{2}}$&$\pm 1$
&$e^{\imath k}$\\ \hline
$_{k}\Upsilon$&2&$A_2$&$\pmatrix{0&e^{\imath k}\cr
1&0\cr}$&$\pmatrix{\imath&0\cr 0&-\imath\cr}$&$e^{\imath k}I_2$\\ \hline
$_{\bf k}G^\pm$&2&$A_2$&$\pmatrix{0&e^{\imath k_x}\\1&0\\}$&
$\pm\pmatrix{e^{\imath\frac{k_y}{2}}&0\\0&e^{-\imath\frac{k_y}{2}}\\}$
&$e^{\imath k_x}I_2$\\ \hline\hline
\end{tabular}}\end{center}\end{table}}}
{\footnotesize{
\begin{table}[hbt]
\caption[]{\label{45}{\footnotesize{
The irrs of the rectangular primitive diperiodic group $\D\g 45$={\bf
C}$_22_1T_h=\gr\{C_2, \zux, \zhy\}$ induced by the element $C_2$ from
the group ${\bf Dg}30$ (Tab. \ref{29}).  The irreducible domain is
presented in the Fig. \ref{PRIMid}$\, (b)$. The quasi angular momentum
$m$ takes on the values $0$ and $1$. Here, $L_2(k)=\pmatrix{0&e^{\imath
k}\cr 1&0\cr}$ and
$K_2=\diag{(e^{\imath\frac{k_y}{2}},e^{-\imath\frac{k_y}{2}})}$.}}}
\begin{center}
{\tabcolsep1truemm  
\begin{tabular}{||c|c||c|c|c||} \hline\hline
Irr&D&$C_2$&$(U_x|\frac{1}{2}\,0)$&$(\gs_h|0\,\frac{1}{2})$\\ \hline \hline
$^v\gG_m^\pm$&1&$(-1)^m$&$(-1)^v$&$\pm 1$\\ \hline
$X^\pm$&2&$A_2$&$\pmatrix{\imath&0\cr 0&-\imath}$&$\pm I_2$\\ \hline
$Y_m$&2&$(-1)^mA_2$&$A_2$&$\pmatrix{\imath&0\cr 0&-\imath}$\\ \hline
$M_m$&2&$(-1)^mA_2$&$\pmatrix{0&-1\cr 1&0\cr}$&
$\pmatrix{\imath&0\cr 0&-\imath}$\\ \hline
$_k^v\Delta^\pm$&2&$A_2$&$(-1)^v\pmatrix{e^{\imath\frac{k}{2}}&0\cr 
0&e^{-\imath\frac{k}{2}}\cr}$&$\pm I_2$\\ \hline
$_k\Lambda_m^\pm$&2&$(-1)^mA_2$&$\pmatrix{0&-1\cr 1&0\cr}$&$
\pm\pmatrix{e^{\imath\frac{k}{2}}&0\cr 0&e^{-\imath\frac{k}{2}}\cr}$\\ \hline
$_k\Phi_m^\pm$&2&$(-1)^mA_2$&$A_2$&$
\pm\pmatrix{e^{\imath k}&0\cr 0&e^{-\imath k}\cr}$\\ \hline
$_k\Upsilon$&4&$\pmatrix{0&I_2\cr I_2&0\cr}$&$
\pmatrix{L_2(k)&0\cr 0&L_2(k)^{-1}\cr}$&$
\pmatrix{\imath&0&0&0\cr 0&-\imath&0&0\cr 0&0&-\imath&0\cr 0&0&0&\imath\cr}$
\\ \hline
$_{\bf k}G^\pm$&4&$\pmatrix{0&I_2\cr I_2&0\cr}$&$
\pmatrix{L_2(k_x)&0\cr 0&L_2(k_x)^{-1}\cr}$&$
\pm\pmatrix{K_2&0\cr 0&K_2^*\cr}$\\ \hline\hline
\end{tabular}}\end{center}\end{table}}}

{\footnotesize{
\begin{table}[hbt]
\caption[]{\label{38}{\footnotesize{
The irrs of the rectangular primitive diperiodic groups 
$\D\g 38={\bf D}_2\T_h=\gr\{C_2, U_y, \tx, \zhy\}$ and 
$\D\g 41={\bf C}_{2v}\T_h=\gr\{C_2, \gs_y, \tx, \zhy\}$ induced by
the element $U_y$ or $\sigma_y$ from the irrs of the group ${\bf Dg}7$
(Tab. \ref{7}).  The irreducible domain is presented in the Fig.
\ref{PRIMid}$\, (b)$. The quasi angular momentum $m$ takes on the
values $0$ and $1$. Here,
$K_2=\diag{(e^{\imath\frac{k_y}{2}},e^{-\imath\frac{k_y}{2}})}$.}}}
\begin{center}
{\tabcolsep1truemm  
\begin{tabular}{||c|c||c|c|c|c||} \hline\hline
Irr&D&$C_2$&$\gs_y$ or $U_y$&$(\gs_h|0\,\frac{1}{2})$&$(I|1\, 0)$
\\ \hline \hline
$^v\gG_m^\pm$&1&$(-1)^m$&$(-1)^v$&$\pm 1$&$1$\\ \hline
$^vX_m^\pm$&1&$(-1)^m$&$(-1)^v$&$\pm 1$&$-1$\\ \hline
$^vY_{(0,1)}$&2&$A_2$&$(-1)^vI_2$&$\pmatrix{\imath&0\cr 0&-\imath}$&
$I_2$\\ \hline
$^vM_{(0,1)}$&2&$A_2$&$(-1)^vI_2$&$\pmatrix{-\imath&0\cr 0&\imath}$&
$-I_2$\\ \hline
$_k^v\gD^\pm$&2&$A_2$&$(-1)^vA_2$&$\pm I_2$&
$\pmatrix{e^{\imath k}&0\\0&e^{-\imath k}\\}$\\ \hline
$_k^v\Phi^\pm$&2&$A_2$&$(-1)^vI_2$&$\pm\pmatrix
{e^{\imath\frac{k}{2}}&0\\0&e^{-\imath\frac{k}{2}}\\}$&$I_2$\\ \hline
$_k^v\Lambda^\pm$&2&$A_2$&$(-1)^vI_2$&$\pm\pmatrix
{e^{\imath\frac{k}{2}}&0\\0&e^{-\imath\frac{k}{2}}\\}$&$-I_2$\\ \hline
$_k\Upsilon$&4&$\pmatrix{A_2&0\\0&A_2\\}$&$\pmatrix{0&I_2\\I_2&0\\}$& 
$\pmatrix{\imath&0&0&0\cr 0&-\imath&0&0\cr 0&0&\imath&0\cr 0&0&0&-\imath\cr}$&
$\pmatrix{e^{\imath k}&0&0\cr 
0&e^{-\imath k}I_2&0\cr 0&0&e^{\imath k}\cr}$\\ \hline
$_{\bf k}G^\pm$&4&$\pmatrix{A_2&0\\0&A_2\\}$&$\pmatrix{0&I_2\\I_2&0\\}$& 
$\pm\pmatrix{K_2&0\\0&K_2\\}$&$\pmatrix{e^{\imath k_x}&0&0\cr 
0&e^{-\imath k_x}I_2&0\cr 0&0&e^{\imath k_x}\cr}$\\ \hline\hline
\end{tabular}}\end{center}\end{table}}}
{\footnotesize{
\begin{table}[hbt]
\caption[]{\label{43}{\footnotesize{
The irrs of the rectangular primitive diperiodic group 
$\D\g 43={\bf D}_12_1T_h=\gr\{U_y, \zux, \zhy\}$ 
induced by the element $U_y$ from the irrs of the group ${\bf Dg}30$
(Tab. \ref{29}).  The irreducible domain is presented in
the Fig. \ref{PRIMid}$\, (b)$. Here, $L_2(k)=\pmatrix{0&e^{\imath k}\cr 1&0\cr}$
and $K_2=\diag{(e^{\imath\frac{k_y}{2}},e^{-\imath\frac{k_y}{2}})}$.}}}
\begin{center}
{\tabcolsep1truemm  
\begin{tabular}{||c|c||c|c|c||} \hline\hline
Irr&D&$U_y$&$(U_x|\frac{1}{2}\,0)$&$(\gs_h|0\, \frac{1}{2})$\\ \hline\hline
$^{\bf v}\gG^\pm$&1&$(-1)^{v_y}$&$(-1)^{v_x}$&$\pm 1$\\ \hline
$X^\pm$&2&$A_2$&$\pmatrix{\imath&0\cr 0&-\imath}$&$\pm I_2$\\ \hline
$^vY$&2&$(-1)^vI_2$&$A_2$&$\pmatrix{\imath&0\cr 0&-\imath}$\\ \hline
$^vM$&2&$(-1)^v\pmatrix{1&0\cr 0&-1\cr}$&$\pmatrix{0&-1\cr 1&0\cr}$&
$\pmatrix{\imath&0\cr 0&-\imath}$\\ \hline
$_k^v\gD^\pm$&2&$A_2$&$(-1)^v\pmatrix{e^{\imath\frac{k}{2}}&0\\
0&e^{-\imath\frac{k}{2}}\\}$&$\pm I_2$\\ \hline
$_k^v\Phi^\pm$&2&$(-1)^vI_2$&$A_2$&$\pm\pmatrix{e^{\imath\frac{k}{2}}&0\\
0&e^{-\imath\frac{k}{2}}\\}$\\ \hline
$_k^v\Lambda^\pm$&2&$(-1)^v\pmatrix{1&0\cr 0&-1\cr}$&$
\pmatrix{0&-1\cr 1&0\cr}$&$\pm\pmatrix{e^{\imath\frac{k}{2}}&0\\
0&e^{-\imath\frac{k}{2}}\\}$\\ \hline
$_k\Upsilon$&4&$\pmatrix{0&I_2\\I_2&0\\}$&
$\pmatrix{L_2(k)&0\cr 0&L_2(k)^{-1}\cr}$&
$\pmatrix{\imath&0&0&0\cr 0&-\imath&0&0\cr 0&0&\imath&0\cr 0&0&0&-\imath\cr}
$\\ \hline
$_{\bf k}G^\pm$&4&$\pmatrix{0&I_2\\I_2&0\\}$&
$\pmatrix{L_2(k_x)&0\cr 0&L_2(k_x)^{-1}\cr}$&
$\pm\pmatrix{K_2&0\cr 0&K_2\cr}$\\ \hline\hline
\end{tabular}}\end{center}\end{table}}}

{\footnotesize{
\begin{table}[hbt]
\caption[]{\label{10}{\footnotesize{
The irrs of the rectangular centered diperiodic groups
$\D\g 10={\bf D}_1\T '=\gr\{U_x, \tx, \zts\}$, 
$\D\g 13={\bf C}_{1v}\T '=\gr\{\gs_x, \tx, \zts\}$,
$\D\g 35={\bf D}_{1h}\T '=\gr\{U_x, \gs_h, \tx, \zts\}$
and the rectangular primitive diperiodic groups
$\D\g 31={\bf D}_1\T_h'=\gr\{U_x, \tx, \zhs\}$,
$\D\g 32={\bf C}_{1v}\T_h'=\gr\{\gs_x, \tx, \zhs\}$,
induced by the elements $(I|\frac{1}{2}\,\frac{1}{2})$ 
and $(\sigma_h|\frac{1}{2}\,\frac{1}{2})$ from the
irrs of the groups ${\bf Dg}8$, ${\bf Dg}11$ and ${\bf Dg}24$ (Tab. \ref{8}).
The irreducible domain is presented in the Fig. \ref{PRIMid}$\, (a)$.}}}
\begin{center}
\begin{tabular}{||c|c||c|c|c|c|c||} \hline \hline 
Irr&D&$\gs_x\ {\rm or}\ U_x$& $\gs_h$&$(I|1\,0)$&$(I|\frac{1}{2}\,
\frac{1}{2})$&$(\sigma_h|\frac{1}{2}\,\frac{1}{2})$\\  \hline\hline
$_{k}^{v,t}\gD^\pm$&1&(-1)$^v$&$\pm 1$&$e^{\imath k}$&$(-1)^t
e^{\imath\frac{k}{2}}$&$\pm e^{\imath\frac{k}{2}}$\\ \hline 
$_k\Upsilon^\pm$&2&$\pmatrix{1&0\cr 0&-1\cr}$&$\pm I_2$&$e^{\imath k}I_2$&
$\pmatrix{0&-e^{\imath k}\\1&0\\}$&$\pmatrix{0&-e^{\imath k}\\1&0\\}$
\\ \hline 
$^t_{\bf k}G^\pm$&2&$A_2$&$\pm I_2$&$e^{\imath k_x}I_2$&
$(-1)^t\pmatrix{e^{\frac{\imath}{2}(k_x+k_y)}&0\cr 
0&e^{\frac{\imath}{2}(k_x- k_y)}\cr}$&
$\pm\pmatrix{e^{\frac{\imath}{2}(k_x+k_y)}&0\cr 
0&e^{\frac{\imath}{2}(k_x- k_y)}\cr}$\\ \hline \hline 
\end{tabular}\end{center}\end{table}}}
{\footnotesize{
\begin{table}[hbt]
\caption[]{\label{16}{\footnotesize{
The irrs of the rectangular centered diperiodic groups 
$\D\g 16={\bf D}_{1d}\T '=\gr\{C_2\gs_h, \gs_x, \tx, \zts\}$, 
$\D\g 22={\bf D}_2\T '=\gr\{C_2, U_x, \tx, \zts\}$, 
$\D\g 34={\bf C}_{2v}\T '=\gr\{C_2, \gs_x, \tx, \zts\}$ and 
$\D\g 47={\bf D}_{2h}\T '=\gr\{C_2, U_x, \gs_h, \tx, \zts\}$ 
induced by the element $(I|\frac{1}{2}\,\frac{1}{2})$ from the
irrs of the groups ${\bf Dg}14$, ${\bf Dg}19$, ${\bf Dg}23$ and 
${\bf Dg}37$ (Tab. \ref{14}). The
irreducible domain is presented in the Fig. \ref{PRIMid}$\, (b)$.
The quasi angular momentum takes on the
values 0 and 1. Here, $K_2=\diag{(e^{\imath k_x},e^{-\imath k_x})}$, 
$L_2=\diag{(e^{\imath k},e^{-\imath k})}$ and 
$K_4=\diag{(e^{\frac{\imath}{2}(k_x+k_y)},e^{-\frac{\imath}{2}(k_x+k_y)},
e^{\frac{\imath}{2}(k_x-k_y)},e^{-\frac{\imath}{2}(k_x-k_y)})}$.}}}
\begin{center}
{\tabcolsep1truemm  
\begin{tabular}{||c|c||c|c|c|c|c||} \hline \hline 
Irr&D&$C_2$ or $C_2\sigma_h$&$\gs_x\ {\rm or}\ U_x$&$\gs_h$& 
$(I|1\,0)$&$(I|\frac{1}{2}\,\frac{1}{2})$\\ \hline\hline
$^{v,t}\gG^\pm_m$&1&(-1)$^m$&$(-1)^v$&$\pm 1$&1&$(-1)^t$\\ \hline 
$^vX_{(0,1)}^\pm$&2&$\pmatrix{1&0\cr 0&-1\cr}$&$(-1)^vI_2$&$\pm I_2$&
$-I_2$&$\pmatrix{0&-1\\1&0\\}$\\ \hline 
$^vY^\pm$&2&$\pmatrix{1&0\cr 0&-1\cr}$&$(-1)^v\pmatrix{1&0\cr 0&-1\cr}$
&$\pm I_2$&$I_2$&$\pmatrix{0&-1\cr 1&0\cr}$\\ \hline 
$M^\pm_m$&2&(-1)$^mI_2$&$\pmatrix{1&0\cr 0&-1\cr}$&$\pm I_2$&$-I_2$&$A_2$\\ 
\hline 
$_k^{v,t}\gD^\pm$&2&$A_2$&$(-1)^vI_2$&$\pm I_2$&
$\pmatrix{e^{\imath k}&0\\0&e^{-\imath k}\\}$&
$(-1)^t\pmatrix{e^{\imath\frac{k}{2}}&0\cr 0&e^{-\imath\frac{k}{2}}\cr}$\\ \hline 
$_k^{v,t}\Phi^\pm$&2&$A_2$&$(-1)^vA_2$&$\pm I_2$&$I_2$&
$(-1)^t\pmatrix{e^{\imath\frac{k}{2}}&0\cr 0&e^{-\imath\frac{k}{2}}\cr}$\\ \hline 
$_{k}\Upsilon^\pm$&4&$\pmatrix{A_2&0&0\\0&0&-e^{-\imath k}\\
0&-e^{\imath k}&0\\}$&$\pmatrix{I_2&0\\0&-I_2\\}$&$\pm I_4$&
$\pmatrix{L_2&0\\0&L_2\\}$&$\pmatrix{0&-L_2\cr I_2&0\cr}$\\ \hline 
$_{k}\Lambda^\pm$&4&$\pmatrix{A_2&0&0\\0&0&-e^{-\imath k}\\
0&-e^{\imath k}&0\\}$&$\pmatrix{A_2&0&0\\0&0&e^{-\imath k}\\
0&e^{\imath k}&0\\}$&$\pm I_4$&$-I_4$&
$\pmatrix{0&-L_2\cr I_2&0\cr}$\\ \hline 
$^t_{\bf k}G^\pm$&4&$\pmatrix{A_2&0\\0&A_2\\}$&
$\pmatrix{0&I_2\\I_2&0\\}$&$\pm I_4$&$\pmatrix{K_2&0\\0&K_2\\}$&
$(-1)^tK_4$\\ \hline \hline 
\end{tabular}}\end{center}\end{table}}}

{\footnotesize{
\begin{table}[hbt]
\caption[]{\label{18}{\footnotesize{
The irrs of the rectangular primitive diperiodic groups 
$\D\g 18={\bf S}_2{\bf T}'_v=\gr\{C_2\gs_h, \tx, (\sigma_y|\frac{1}{2}\,\frac{1}{2})\}$, 
$\D\g 21={\bf C}_2{\bf 2}'_1=\gr\{C_2, \tx, (U_y|\frac{1}{2}\,\frac{1}{2})\}$,
$\D\g 33={\bf C}_2{\bf T}'_v=\gr\{C_2, \tx, (\sigma_y|\frac{1}{2}\,\frac{1}{2})\}$ and 
$\D\g 44={\bf C}_{2h}\T_v'=\gr\{C_2, \gs_h, \tx, (\sigma_y|\frac{1}{2}\,\frac{1}{2})\}$
induced by the elements $(\gs_y |\frac{1}{2}\,\frac{1}{2})$ and
$(U_y|\frac{1}{2}\,\frac{1}{2})$ from the 
irrs of the groups ${\bf Dg}2$, ${\bf Dg}3$ and ${\bf Dg}6$ 
(Tab. \ref{2}). The irreducible domain is presented in the 
Fig. \ref{PRIMid}$\, (b)$.
The quasi angular momentum $m$ takes on the values 0 and 1.}}} 
\begin{center}
{\tabcolsep1truemm  
\begin{tabular}{||c|c||c|c|c|c||} \hline\hline
Irr&D&$C_2$ or $C_2\sigma_h$&$\gs_h$&$(\gs_y|\frac{1}{2}\,\frac{1}{2})\ 
{\rm or}\ (U_y|\frac{1}{2}\,\frac{1}{2})$&$(I|1\, 0)$\\ \hline \hline
$^v\gG_m^\pm$&1&$(-1)^m$&$\pm 1$&$(-1)^v$&$1$\\ \hline
$^vM_m^\pm$&1&$(-1)^m$&$\pm 1$&$(-1)^v$&$-1$\\ \hline
$X^\pm$&2&$\pmatrix{1&0\\0&-1\\}$&$\pm I_2$&$A_2$&$-I_2$\\ \hline
$Y^\pm$&2&$\pmatrix{1&0\\0&-1\\}$&$\pm I_2$&$\pmatrix{0&-1\\1&0\\}$&
$I_2$\\ \hline
$^v_k\Delta^\pm$&2&$A_2$&$\pm I_2$&$(-1)^v
\pmatrix{0&e^{-\imath\frac{k}{2}}\cr e^{\imath\frac{k}{2}}&0\cr}$&
$\pmatrix{e^{\imath k_x}&0\cr 0&e^{-\imath k_x}\cr}$\\ \hline
$^v_k\Upsilon^\pm$&2&$A_2$&$\pm I_2$&$(-1)^v
\pmatrix{0&e^{-\imath\frac{k}{2}}\cr -e^{\imath\frac{k}{2}}&0\cr}$&
$\pmatrix{e^{\imath k_x}&0\cr 0&e^{-\imath k_x}\cr}$\\ \hline
$^v_k\Phi^\pm$&2&$A_2$&$\pm I_2$&$(-1)^v\pmatrix{e^{\imath\frac{k}{2}}&0\cr 
0&e^{-\imath\frac{k}{2}}\cr}$&$I_2$\\ \hline
$^v_k\Lambda^\pm$&2&$A_2$&$\pm I_2$&$(-1)^v\pmatrix{e^{\imath\frac{k}{2}}&0\cr 
0&-e^{-\imath\frac{k}{2}}\cr}$&$-I_2$\\ \hline
$_{\bf k}G^\pm$&4&$\pmatrix{A_2&0&0\cr 0&0&e^{\imath (k_x-k_y)}\cr
0&e^{-\imath (k_x-k_y)}&0\cr}$&$\pm I_4$& 
$\pmatrix{0&e^{\imath k_y}&0\cr 0&0&e^{-\imath k_y}\cr 
I_2&0&0\cr}$&$\pmatrix{e^{\imath k_x}&0&0\\0&e^{-\imath k_x}I_2&0\\
0&0&e^{\imath k_x}}$\\ \hline\hline
\end{tabular}}\end{center}\end{table}}}

{\footnotesize{
\begin{table}[hbt]
\caption[]{\label{36}{\footnotesize{
The irrs of the rectangular centered diperiodic group $\D\g 36={\bf
D}_1{\bf T}_h=\gr\{U_y, \zts, \zhy\}$ induced by the elements $U_y$
and $(I|\frac{1}{2}\,\frac{1}{2})$ from the irrs of the group ${\bf
Dg}5$ (Tab. \ref{1}).  The irreducible domain is presented in the Fig.
\ref{PRIMid}$\, (c)$.}}}
\begin{center}
{\tabcolsep1truemm  
\begin{tabular}{||c|c||c|c|c||} \hline\hline
Irr&D&$U_y$&$(\gs_h|0\, \frac{1}{2})$&$(I|\frac{1}{2}\,\frac{1}{2})$\\ 
\hline \hline
$_k^{v,t}\Phi^\pm$&1&$(-1)^v$&$\pm e^{\imath\frac{k_y}{2}}$&
$(-1)^te^{\imath\frac{k}{2}}$\\ \hline
$_k\Lambda^\pm$&2&$\pmatrix{1&0\cr 0&-1\cr}$&$\pm e^{\imath\frac{k}{2}}I_2$&
$\pmatrix{0&-e^{\imath k}\\1&0\\}$\\ \hline
$_{\bf k}^tG^\pm$&2&$A_2$&$\pm e^{\imath\frac{k_y}{2}}I_2$&
$(-1)^t
\pmatrix{e^{\frac{\imath}{2}(k_x+k_y)}&0\\0&e^{-\frac{\imath}{2}(k_x-k_y)}\\}$
\\ \hline\hline
\end{tabular}}\end{center}\end{table}}}

{\footnotesize{
\begin{table}[hbt]
\caption[]{\label{39}{\footnotesize{
The irrs of the rectangular primitive diperiodic groups 
$\D\g 39={\bf D}_2\T_h'=\gr\{C_2, U_x, \tx, \zhs\}$, 
$\D\g 42={\bf D}_{1d}\T_h'=\gr\{C_2\gs_h, U_x, \tx, \zhs\}$ and
$\D\g 46={\bf C}_{2v}\T_h'=\gr\{C_2, \gs_x, \tx, \zhs\}$ 
induced by the element 
$(\gs_h|\frac{1}{2}\,\frac{1}{2})$ from the irrs of the groups 
${\bf Dg}19$, ${\bf Dg}14$ and ${\bf Dg}23$, respectively (Tab. \ref{14}). 
The irreducible domain is presented in the Fig. \ref{PRIMid}$\, (b)$.
The angular momentum $m$ takes on the values 0 and 1. Here,
$K_2=\diag{(e^{\imath\frac{k}{2}},e^{-\imath\frac{k}{2}})}$,
$L_2(k)=\diag{(e^{\imath k},e^{-\imath k})}$ and
$K_4=\diag{(e^{\frac{\imath}{2}(k_x+k_y)},e^{-\frac{\imath}{2}(k_x+k_y)},
e^{\frac{\imath}{2}(k_x-k_y)},e^{-\frac{\imath}{2}(k_x-k_y)})}$.}}}
\begin{center}
{\tabcolsep1truemm  
\begin{tabular}{||c|c||c|c|c|c||} \hline\hline
Irr&D&$C_2$ or $C_2\sigma_h$&$\gs_x\ {\rm or}\ U_x$& 
$(\gs_h|\frac{1}{2}\,\frac{1}{2})$&$(I|1\, 0)$\\ \hline \hline
$^v\gG_m^\pm$&1&$(-1)^m$&$(-1)^v$&$\pm 1$&1\\ \hline
$^vX_{(1,0)}$&2&$\pmatrix{1&0\cr 0&-1\cr}$&$(-1)^vI_2$&
$\pmatrix{0&-1\cr 1&0\cr}$&$-I_2$\\ \hline
$^vY$&2&$\pmatrix{1&0\cr 0&-1\cr}$&$(-1)^v\pmatrix{1&0\cr 0&-1\cr}$&
$\pmatrix{0&-1\cr 1&0\cr}$&$I_2$\\ \hline
$M_m$&2&$(-1)^mI_2$&$\pmatrix{1&0\cr 0&-1\cr}$&$A_2$&$-I_2$\\ \hline
$_k^v\Delta^\pm$&2&$A_2$&$(-1)^vI_2$&$\pm K_2$&$\pmatrix{e^{\imath k}&0\cr 
0&e^{-\imath k}\cr}$\\ \hline
$_k^v\Phi^\pm$&2&$A_2$&$(-1)^vA_2$&$\pm K_2$&$I_2$\\ \hline
$_k\Upsilon$&4&$\pmatrix{A_2&0&0\cr 0&0&-e^{-\imath k}\cr 
0&-e^{\imath k}&0\cr}$&$\pmatrix{I_2&0\cr 0&-I_2\cr}$&
$\pmatrix{0&-L_2(k)\cr I_2&0\cr}$&$\pmatrix{L_2(k)&0\cr 0&L_2(k)\cr}$\\ \hline
$_k\Lambda$&4&$\pmatrix{A_2&0&0\cr 0&0&-e^{-\imath k}\cr 
0&-e^{\imath k}&0\cr}$&$\pmatrix{A_2&0&0\cr 0&0&e^{-\imath k}\cr 
0&e^{\imath k}&0\cr}$&$\pmatrix{0&-L_2(k)\cr I_2&0\cr}$&$-I_4$\\ \hline
$_{\bf k}G^\pm$&4&$\pmatrix{A_2&0\\0&A_2\\}$&$\pmatrix{0&I_2\\I_2&0\\}$& 
$\pm K_4$&$\pmatrix{L_2(k_x)&0\cr 0&L_2(k_x)\cr}$\\ \hline\hline
\end{tabular}}\end{center}\end{table}}}
{\footnotesize{
\begin{table}[hbt]
\caption[]{\label{48}{\footnotesize{
The irrs of the rectangular centered diperiodic group $\D\g 48={\bf
C}_{2v}{\bf T}_h'=\gr\{C_2, \gs_y, \zts, \zhy\}$ induced by the
element $(I|\frac{1}{2}\,\frac{1}{2})$ from the irrs of the group ${\bf
Dg}41$ (Tab. \ref{38}).  The irreducible domain is presented in the
Fig. \ref{PRIMid}$\, (b)$.  The angular momentum $m$ takes on the values 0 and
1. Here,
$K_2(k)=\diag{(e^{\imath\frac{k}{2}},e^{-\imath\frac{k}{2}})}$,
$L_2=\diag{(e^{\imath k},e^{-\imath k})}$,
$K_4=\diag{(e^{\frac{\imath}{2}(k_x+k_y)},e^{-\frac{\imath}{2}(k_x+k_y)},
e^{-\frac{\imath}{2}(k_x-k_y)},e^{\frac{\imath}{2}(k_x-k_y)})}$ and
$L_4=\diag{(\imath e^{\frac{\imath}{2}k}, -\imath
e^{-\frac{\imath}{2}k},\imath e^{-\frac{\imath}{2}k}, -\imath
e^{\frac{\imath}{2}k})}$.}}}
\begin{center}
{\tabcolsep1truemm  
\begin{tabular}{||c|c||c|c|c|c||} \hline\hline
Irr&D&$C_2$&$\gs_y$&$(\gs_h|0\,\frac{1}{2})$&$(I|\frac{1}{2}\,\frac{1}{2})$
\\ \hline\hline
$^{v,t}\gG_m^\pm$&1&$(-1)^m$&$(-1)^v$&$\pm 1$&$(-1)^t$\\ \hline
$X_m^\pm$&2&$(-1)^m\pmatrix{1&0\cr 0&-1\cr}$&
$\pmatrix{1&0\cr 0&-1\cr}$&$\pm I_2$&$\pmatrix{0&-1\cr 1&0\cr}$\\ \hline
$^{v,t}Y_{(0,1)}$&2&$A_2$&$(-1)^vI_2$&$\pmatrix{\imath&0\cr 0&-\imath\cr}$&
$(-1)^t\pmatrix{\imath&0\cr 0&-\imath\cr}$\\ \hline
$M_{(0,1)}$&4&$\pmatrix{A_2&0\cr 0&A_2\cr}$&$\pmatrix{I_2&0\cr 0&-I_2\cr}$&
$\pmatrix{-\imath&0&0&0\cr 0&\imath&0&0\cr 0&0&-\imath&0\cr 0&0&0&\imath\cr}$&
$\pmatrix{0&I_2\cr I_2&0\cr}$\\ \hline
$_k^{v,t}\Delta^\pm$&2&$A_2$&$(-1)^vA_2$&$\pm I_2$&$(-1)^tK_2(k)$\\ \hline
$_k^{v,t}\Phi^\pm$&2&$A_2$&$(-1)^vI_2$&$\pm K_2(k)$&
$(-1)^tK_2(k)$\\ \hline
$_k\Lambda^\pm$&4&$\pmatrix{A_2&0&0\\0&0&-e^{-\imath k}\\0&-e^{\imath k}&0\\}
$&$\pmatrix{I_2&0\\0&-I_2\\}$&$\pm\pmatrix{K_2(k)&0\cr 0&K_2(k)\cr}$&
$\pmatrix{0&-L_2\cr I_2&0\cr}$\\ \hline
$_k^t\Upsilon$&4&$\pmatrix{A_2&0\\0&A_2\\}$& 
$\pmatrix{0&I_2\\I_2&0\\}$&
$\pmatrix{-\imath&0&0&0\cr 0&\imath&0&0\cr 0&0&-\imath&0\cr 0&0&0&\imath\cr}$&
$(-1)^tL_4$\\ \hline
$_{\bf k}^tG^\pm$&4&$\pmatrix{A_2&0\\0&A_2\\}$& 
$\pmatrix{0&I_2\\I_2&0\\}$&$\pm\pmatrix{K_2(k_y)&0\cr 0&K_2(k_y)\cr}$&
$(-1)^tK_4$\\ \hline\hline
\end{tabular}}\end{center}\end{table}}}

\clearpage

\subsection{The irrs of the square groups}
The irrs of the square diperiodic groups ${\bf Dg}49$ - ${\bf Dg}64$
are presented in the tables \ref{49}-\ref{62} and the corresponding
irreducible domains of the {\Brillouin} zone are given in the Fig.
\ref{SQUAREid}.

{\footnotesize
\begin{figure}[hbt]\begin{center}
\unitlength=0.80mm
\linethickness{0.4pt}
\begin{picture}(120.00,70.00)

\put(30.00,10.00){\makebox(0,0)[cc]{$(a)$}}
\linethickness{1pt}
\put(50.00,60.00){\line(0,-1){20.00}}
\put(50.00,40.00){\line(-1,0){20.00}}
\linethickness{0.1pt}
\put(30.00,40.00){\vector(1,0){30.00}}
\put(30.00,40.00){\vector(0,1){30.00}}
\put(25.00,68.00){\makebox(0,0)[cc]{$k_2$}}
\put(59.00,36.00){\makebox(0,0)[cc]{$k_1$}}
\put(30.00,40.00){\line(0,1){20.00}}
\put(10.00,20.00){\line(1,0){40.00}}
\put(10.00,20.00){\line(0,1){40.00}}
\put(50.00,20.00){\line(0,1){20.00}}
\put(10.00,60.00){\line(1,0){40.00}}
\put(30.00,42.00){\line(1,0){20.00}}
\put(30.00,44.00){\line(1,0){20.00}}
\put(30.00,46.00){\line(1,0){20.00}}
\put(30.00,48.00){\line(1,0){20.00}}
\put(30.00,50.00){\line(1,0){7.00}}
\put(42.00,50.00){\line(1,0){8.00}}
\put(30.00,52.00){\line(1,0){7.00}}
\put(42.00,52.00){\line(1,0){8.00}}
\put(30.00,54.00){\line(1,0){7.00}}
\put(42.00,54.00){\line(1,0){8.00}}
\put(30.00,56.00){\line(1,0){20.00}}
\put(30.00,58.00){\line(1,0){20.00}}

\put(50.00,60.00){\circle*{2.00}}
\put(50.00,40.00){\circle*{2.00}}
\put(30.00,40.00){\circle*{2.00}}
\put(50.00,64.00){\makebox(0,0)[cc]{$M$}}
\put(55.00,41.00){\makebox(0,0)[cb]{$X$}}
\put(30.00,35.00){\makebox(0,0)[cc]{$\Gamma$}}
\put(39.00,52.00){\makebox(0,0)[cc]{$G$}}
\put(90.00,10.00){\makebox(0,0)[cc]{$(b)$}}
\linethickness{1pt}
\put(110.00,60.00){\line(-1,-1){20.00}}
\put(110.00,60.00){\line(0,-1){20.00}}
\put(110.00,40.00){\line(-1,0){20.00}}
\linethickness{0.1pt}
\put(90.00,40.00){\vector(1,0){30.00}}
\put(90.00,40.00){\vector(0,1){30.00}}
\put(85.00,68.00){\makebox(0,0)[cc]{$k_2$}}
\put(119.00,36.00){\makebox(0,0)[cc]{$k_1$}}

\put(70.00,20.00){\line(1,0){40.00}}
\put(70.00,20.00){\line(0,1){40.00}}
\put(110.00,20.00){\line(0,1){20.00}}
\put(70.00,60.00){\line(1,0){40.00}}
\put(92.00,42.00){\line(1,0){18.00}}

\put(94.00,44.00){\line(1,0){8.00}}
\put(108.00,44.00){\line(1,0){2.00}}
\put(96.00,46.00){\line(1,0){6.00}}
\put(108.00,46.00){\line(1,0){2.00}}
\put(98.00,48.00){\line(1,0){4.00}}
\put(108.00,48.00){\line(1,0){2.00}}
\put(100.00,50.00){\line(1,0){10.00}}
\put(102.00,52.00){\line(1,0){8.00}}
\put(104.00,54.00){\line(1,0){6.00}}
\put(106.00,56.00){\line(1,0){4.00}}
\put(108.00,58.00){\line(1,0){2.00}}

\put(110.00,60.00){\circle*{2.00}}
\put(110.00,40.00){\circle*{2.00}}
\put(90.00,40.00){\circle*{2.00}}
\put(110.00,64.00){\makebox(0,0)[cc]{$M$}}
\put(115.00,41.00){\makebox(0,0)[cb]{$X$}}
\put(90.00,35.00){\makebox(0,0)[cc]{$\Gamma$}}
\put(105.00,46.00){\makebox(0,0)[cc]{$G$}}
\end{picture}
\caption[]{\label{SQUAREid}{\footnotesize 
The irreducible domains of the {\Brillouin} zone for the square
diperiodic groups. The special points (filled circles) are $\gG
=(0,0)$, $X=(\pi ,0)$ and $M=(\pi ,\pi )$, while the special lines are
$\gD =(k,0)$, $\gL =(\pi ,k)$ and $\gS =(k,k)$, with $k\in (0,\pi )$.}}
\end{center}\end{figure}
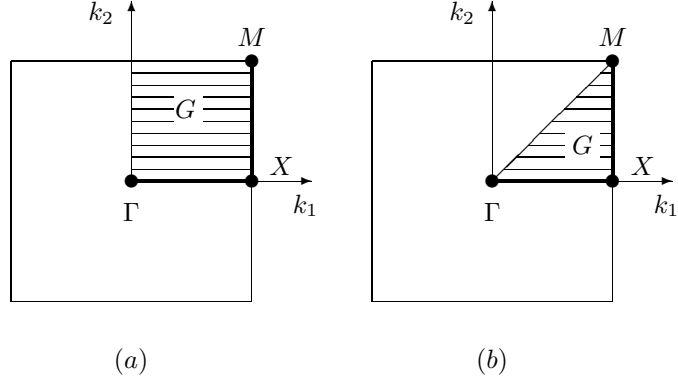}

The following notation is used: $B_4=\pmatrix{0&1\cr I_3&0\cr}$,
$C_2=\diag{(-1,e^{\imath k})}$, 
$D_2=\diag{(e^{\imath k},-1)}$, $D_4=\pmatrix{0&0&0&-1\cr e^{\imath
k}&0&0&0\cr 0&-1&0&0\cr 0&0&e^{-\imath k}&0\cr}$,
$E_2=\diag{(e^{\imath (k_x+k_y)/2},e^{\imath (k_x-k_y)/2})}$,
$F_2=\diag{(e^{\imath\frac{k}{2}},e^{-\imath\frac{k}{2}})}$,
$J_2=\diag{(e^{\imath k},e^{-\imath k})}$, $K_2=\diag{(e^{\imath
k_x},e^{-\imath k_y})}$, $L_2=\diag{(e^{\imath k_y},e^{\imath k_x})}$,
$P_2=\diag{(e^{\imath k},1)}$ and 
$R_2=\diag{(1,e^{\imath k})}$.  
{\footnotesize{
\begin{table}[hbt]
\caption[]{\label{49}{\footnotesize{The irrs of the square diperiodic groups 
$\D\g 49={\bf C}_4\T =\gr\{C_4, \tx, \ty\}$, 
$\D\g 50={\bf S}_4\T =\gr\{C_4\gs_h, \tx, \ty\}$ and 
$\D\g 51={\bf C}_{4h}\T =\gr\{C_4, \gs_h,\tx, \ty\}$
induced by the elements of the groups ${\bf C}_4$ and ${\bf S}_4$ from
the irrs of the groups ${\bf Dg}1$ and ${\bf Dg}4$ (Tab.
\ref{1}). For the $\gG^\pm_m$ and $\M^\pm_m$ the quasi angular momentum
$m$ takes on the integer values form the interval $[-1,2]$, while for
the $\X^\pm_m$ it takes on the values $0$ and $1$, only.  The
irreducible domain of the {\Brillouin} zone is presented in the Fig.
\ref{SQUAREid}$\, (a)$.}}}
\begin{center}
\begin{tabular}{||c|c||c|c|c|c||} \hline \hline 
Irr&D&$C_4$ or $C_4\sigma_h$&$\gs_h$&$(I|1\,0)$&$(I|0\,1)$\\ \hline\hline
$\gG_m^{\pm}$ &1& $\imath ^m$ &$\pm 1$& 1 & 1   \\ \hline 
$\M_m^{\pm}$& 1 & $\imath ^m$ & $\pm 1$& -1 & -1  \\ \hline 
$\X_m^{\pm}$&2&$\pmatrix{0&(-1)^m\cr 1&0\cr}$&$\pm I_2$&
$\pmatrix{-1&0\\0&1\\}$&$\pmatrix{1&0\\0&-1\\}$ \\ \hline 
$_{\bf k}G^{\pm}$&4&$B_4$&$\pm I_4$&
$\pmatrix{K_2&0\\0&K_2^*\\}$&$\pmatrix{L_2&0\\0&L_2^*\\}$\\ \hline \hline 
\end{tabular}\end{center}\end{table}}}
{\footnotesize{
\begin{table}[hbt]
\caption[]{\label{52}{\footnotesize{
The irrs of the square diperiodic group 
$\D\g 52={\bf C}_4\T_h'=\gr\{C_4, \tx, \zhs\}$ induced by the element 
$(\gs_h|\frac{1}{2}\,\frac{1}{2})$ 
from the irrs of the group ${\bf Dg}49$
(Tab. \ref{49}). For $\gG^\pm_m$ the quasi angular 
momentum $m$ takes on the values -1,0,1 and 2, and for $M_m$ the possible
values are 0 and 1.
The irreducible domain of the {\Brillouin} zone is presented in the 
Fig. \ref{SQUAREid}$\, (a)$.}}} 
\begin{center}
{\tabcolsep1truemm  
\begin{tabular}{||c|c||c|c|c||} \hline\hline
Irr&D&$C_4$&$(\gs_h|\frac{1}{2} \,\frac{1}{2})$&$(I|1\, 0)$\\ \hline\hline
$\gG_m^\pm$&1&$\imath^m$&$\pm 1$&$1$\\ \hline
$M_m$&2&$\imath^m\pmatrix{1&0\cr 0&-1\cr}$&$A_2$&$-I_2$\\ \hline
$X$&4&$\pmatrix{A_2&0&0\cr 0&0&-1\cr 0&1&0\cr}$&
$\pmatrix{0&-I_2\cr I_2&0\cr}$&
$\pmatrix{-1&0&0&0\cr 0&1&0&0\cr 0&0&-1&0\cr 0&0&0&1\cr}$\\ \hline 
$_{\bf k}G^\pm$&4&$B_4$&$\pm\pmatrix{E_2&0\cr 0&E_2^*\cr}$&
$\pmatrix{K_2&0\cr 0&K_2^*\cr}$\\ \hline\hline
\end{tabular}}\end{center}\end{table}}}
{\footnotesize{
\begin{table}[hbt]
\caption[]{\label{53}{\footnotesize{
The irrs of the square diperiodic groups 
$\D\g 53={\bf D}_4\T=\gr\{C_4, U_x, \tx, \ty\}$, 
$\D\g 55={\bf C}_{4v}\T=\gr\{C_4, \gs_x,\tx, \ty\}$, 
$\D\g 57={\bf D}_{2d}\T =\gr\{C_4\gs_h, U_x, \tx, \ty\}$, 
$\D\g 59={\bf D}_{2d}\T=\gr\{C_4\gs_h, \gs_x, \tx, \ty\}$ and
$\D\g 61={\bf D}_{4h}\T =\gr\{C_4, \gs_x, \gs_h, \tx, \ty\}$ induced
by the element $U_x$ and $\sigma_x$ from the irrs of the groups ${\bf
Dg}49$, ${\bf Dg}50$ and ${\bf Dg}51$ (Tab. \ref{49}).  For 
$^v\gG^\pm_m$ and $^v\M^\pm_m$ the quasi angular momentum $m$ takes on
the values $0$ and $2$, while for the $^vX^\pm_m$ it takes on the
values $0$ and $1$.  The irreducible domain of the {\Brillouin} zone is
presented in the Fig. \ref{SQUAREid}$\, (b)$.}}}
\begin{center}
{\tabcolsep1truemm  
\begin{tabular}{||c|c||c|c|c|c|c||} \hline \hline
Irr&D&$C_4$ or $C_4\sigma_h$&$\gs_x$ or $U_x$ &$\gs_h$&$(I|1\,0)$&$(I|0\,1)$\\ 
\hline\hline
$^v\gG^\pm_m$&1&$\imath^m$&$(-1)^v$&$\pm 1$&$1$&$1$\\ \hline
$^v\M^\pm_m$&1&$\imath^m$&$(-1)^v$&$\pm 1$&$-1$&$-1$\\ \hline
$\gG_1^\pm$&2&$\pmatrix{\imath&0\cr 0&-\imath\cr}$&
$\pmatrix{0&1\cr 1&0\cr}$&$\pm I_2$&$I_2$&$I_2$\\ \hline 
$\M_1^\pm$&2&$\pmatrix{\imath&0\cr 0&-\imath\cr}$&
$\pmatrix{0&1\cr 1&0\cr}$&$\pm I_2$&$-I_2$&$-I_2$\\ \hline 
$^v\X^\pm_m$&2&$\pmatrix{0&(-1)^m\cr 1&0\cr}$&$(-1)^v
\pmatrix{1&0\cr 0&(-1)^m\cr}$&
$\pm I_2$&$\pmatrix{-1&0\cr0&1\cr}$&$\pmatrix{1&0\cr0&-1\cr}$\\ \hline 
$^v_k\Sigma^\pm$&4&$B_4$&$(-1)^vA_4$&$\pm I_4$&
$\pmatrix{J_2&0\cr 0&J_2^*\cr}$&
$\pmatrix{e^{\imath k}I_2&0\cr 0&e^{-\imath k}I_2\cr}$\\ \hline 
$^v_k\gD^\pm$&4&$B_4$&$(-1)^v\pmatrix{1&0\cr 0&A_3\cr}$&$\pm I_4$&
$\pmatrix{P_2&0\cr 0&P_2^*\cr}$&
$\pmatrix{R_2&0\cr 0&R_2^*\cr}$\\ \hline 
$^v_k\Lambda^\pm$&4&$B_4$&$(-1)^v\pmatrix{A_3&0\cr 0&1\cr}$&$\pm I_4$&
$\pmatrix{C_2^*&0\cr 0&C_2\cr}$&$\pmatrix{D_2&0\cr 0&D_2^*\cr}$\\ \hline 
$_{\bf k}G^\pm$&8&$\pmatrix{B_4&0\cr 0&B_4^{-1}\cr}$&
$\pmatrix{0&I_4\cr I_4&0\cr}$&$\pm I_8$&$
\pmatrix{K_2&0&0&0\cr 0&K_2^*&0&0\cr 0&0&K_2&0\cr 0&0&0&K_2^*\cr}$&
$\pmatrix{L_2&0&0&0\cr 0&L_2^*&0&0\cr 0&0&L_2^*&0\cr 0&0&0&L_2\cr}$\\  
\hline \hline
\end{tabular}}\end{center}\end{table}}}

{\footnotesize{
\begin{table}[hbt]
\caption[]{\label{54}{\footnotesize{
The irrs of the square diperiodic groups  
$\D\g 54={\bf C}_4{\bf 2}'_1=\gr\{C_4, \tx, \zus\}$, 
$\D\g 56={\bf C}_4\T'_v=\gr\{C_4, \tx, \zvs\}$,
$\D\g 58={\bf S}_4{\bf 2}'_1=\gr\{C_4\gs_h, \tx, \zus\}$, 
$\D\g 60={\bf S}_4\T'_v=\gr\{C_4\gs_h, \tx, \zvs\}$ and 
$\D\g 63={\bf C}_{4h}\T_v'=\gr\{C_4, \gs_h, \tx, \zvs\}$ 
induced by the elements 
$(\gs |\frac{1}{2}\,\frac{1}{2})$ and $(U|\frac{1}{2}\,\frac{1}{2})$ from
the irrs of the groups ${\bf Dg}49$, ${\bf Dg}50$ and ${\bf Dg}51$ 
(Tab. \ref{49}).  For the irr $^v\gG_m^\pm$ the quasi angular momentum
$m$ takes on the values 0 and 2, while for the irr $^vM_m^\pm$ it takes
on the values 1 and $-1$. The irreducible domain of the {\Brillouin}
zone is presented in the Fig. \ref{SQUAREid}$\, (b)$. Here,
$H_2=\diag{(e^{\imath (k_x+k_y)},e^{\imath (k_x-k_y)})}$,
$O_2=\diag{(e^{\imath k} ,e^{-\imath k})}$, $S_2=\diag{(e^{\imath
k},1)}$, $N_2=\pmatrix{0&e^{\imath\frac{k}{2}}\cr
e^{\imath\frac{k}{2}}&0\cr}$, and $C_4=\pmatrix{0&e^{-\imath
k_y}&0&0\cr 0&0&e^{-\imath k_x}&0\cr 0&0&0&e^{\imath k_y}\cr e^{\imath
k_x}&0&0&0\cr}$.  }}}
\begin{center}
{\tabcolsep1truemm  
\begin{tabular}{||c|c||c|c|c|c||} \hline\hline
Irr&D&$C_4$ or $C_4\sigma_h$&$\gs_h$&$(\gs|\frac{1}{2}\,\frac{1}{2})$ or 
$(U|\frac{1}{2}\,\frac{1}{2})$&$(I|1\, 0)$\\ \hline \hline
$^v\gG_m^\pm$&1&$\imath^m$&$\pm 1$&$(-1)^v$&$1$ \\ \hline
$^vM_m^\pm$&1&$\imath^m$&$\pm 1$&$(-1)^v$&$-1$ \\ \hline
$\gG_1^\pm$&2&$\pmatrix{\imath&0\cr 0&-\imath\cr}$&$\pm I_2$&$A_2$&$I_2$
\\ \hline
$M_0^\pm$&2&$\pmatrix{1&0\cr 0&-1\cr}$&$\pm I_2$&
$A_2$&$-I_2$\\ \hline
$X^\pm$&4&$\pmatrix{A_2&0&0\cr 0&0&1\cr 0&-1&0\cr}$&
$\pm I_4$&$\pmatrix{0&-I_2\cr I_2&0\cr}$&$\pmatrix{-1&0&0\cr 0&I_2&0\cr
0&0&-1\cr}$\\ \hline
$^v_k\Sigma^\pm$&4&$B_4$&$\pm I_4$&
$(-1)^v\pmatrix{e^{\imath k}&0&0&0\cr 0&0&0&1\cr 
0&0&e^{-\imath k}&0\cr 0&1&0&0\cr}$&
$\pmatrix{O_2&0\cr 0&O_2^*\cr}$\\ \hline
$_k^v\Delta^\pm$&4&$B_4$&$\pm I_4$&
$(-1)^v\pmatrix{N_2&0\cr 0&N_2^*\cr}$&$\pmatrix{S_2&0\cr 0&S_2^*\cr}$\\ \hline
$_k^v\Lambda^\pm$&4&$B_4$&$\pm I_4$&
$(-1)^v\pmatrix{0&0&0&e^{\imath\frac{k}{2}}\cr 
0&0&e^{-\imath\frac{k}{2}}&0\cr 0&-e^{-\imath\frac{k}{2}}&0&0\cr 
-e^{\imath\frac{k}{2}}&0&0&0\cr}$&$\pmatrix{C_2^*&0\cr 0&C_2\cr}$\\ \hline
$_{\bf k}G^\pm$&8&$\pmatrix{B_4&0\cr 0&C_4\cr}$&$\pm I_8$&
$\pmatrix{0&0&H_2&0\cr 0&0&0&H_2^*\cr I_2&0&0&0\cr 0&I_2&0&0\cr}$&
$\pmatrix{K_2&0&0&0\cr 0&K_2^*&0&0\cr 0&0&L_2&0\cr 0&0&0&L_2^*\cr}$\\ 
\hline\hline
\end{tabular}}\end{center}\end{table}}}

{\footnotesize{
\begin{table}[hbt]
\caption[]{\label{62}{\footnotesize{The irrs of the square diperiodic
groups $\D\g 62$={\bf D}$_{2d}\T'_h=\gr\{C_4\gs_h, U_x, \tx, \zhs\}$ and
$\D\g 64$={\bf D}$_{2d}\T'_h=\gr\{C_4\gs_h, \gs_x, \tx, \zhs\}$
induced by the element $(\gs_h|\frac{1}{2}\,\frac{1}{2})$ from the irrs
of the groups ${\bf Dg}57$ and ${\bf Dg}59$ (Tab. \ref{53}).  The quasi
angular momentum $m$ takes on the values 0 and 2.  The irreducible
domain of the {\Brillouin} zone is presented in the Fig.
\ref{SQUAREid}$\, (b)$.  Here, $F_2=\diag{(e^{\imath
(k_x-k_y)/2},e^{\imath (-k_x-k_y)/2})}$ and
$E_4=\pmatrix{0&0&e^{-\imath k}&0\cr 0&-1&0&0\cr e^{\imath k}&0&0&0\cr
0&0&0&-1\cr}$.}}}
\begin{center}
{\tabcolsep1truemm  
\begin{tabular}{||c|c||c|c|c|c||} \hline\hline
Irr&D&$C_4\sigma_h$&$U_x$ or $\gs_x$&$(\gs_h|\frac{1}{2}\, \frac{1}{2})$&$(I|1\, 0)$
\\ \hline \hline
$^v\gG^\pm_m$&1&$\imath^m$&$(-1)^v$&$\pm 1$&1 \\ \hline
$\Gamma_1^\pm$&2&$\pmatrix{\imath&0\\0&-\imath\\}$&$A_2$
&$\pm I_2$&$I_2$\\ \hline
$^vM_0$&2&$(-1)^v\pmatrix{1&0\\0&-1\\}$&$\pmatrix{1&0\\0&-1\\}$
&$A_2$&$-I_2$\\ \hline
$M_1^\pm$&2&$\pmatrix{\imath&0\\0&-\imath\\}$&$A_2$&
$\pm\pmatrix{0&\imath\cr -\imath&0\cr}$&$-I_2$\\ \hline
$^vX$&4&$\pmatrix{A_2&0&0\cr 0&0&-1\cr 0&1&0\cr}$&
$(-1)^v\pmatrix{I_3&0\cr 0&-1\cr}$&$\pmatrix{0&-I_2\cr I_2&0\cr}$&
$\pmatrix{-1&0&0&0\cr 0&1&0&0\cr 0&0&-1&0\cr 0&0&0&1\cr}$\\ \hline
$^v_k\Sigma^\pm$&4&$B_4$&$(-1)^vA_4$&$\pm
\pmatrix{P_2&0\cr 0&P_2^*\cr}$&$\pmatrix{J_2&0\cr 0&J_2^*\cr}$\\ \hline
$^v_k\Delta^\pm$&4&$B_4$&$(-1)^v
\pmatrix{1&0\cr 0&A_3\cr}$&$\pm
\pmatrix{e^{\imath\frac{k}{2}}I_2&0\cr 0&e^{-\imath\frac{k}{2}}I_2\cr}$&
$\pmatrix{P_2&0\cr 0&P_2^*\cr}$\\ \hline
$_k\Lambda$&8&$\pmatrix{B_4&0\cr 0&D_4\cr}$&$\pmatrix{A_3&0&0\cr 
0&1&0\cr 0&0&E_4\cr}$&
$\pmatrix{0&-e^{\imath k}&0&0\cr 0&0&-e^{-\imath k}I_2&0\cr
0&0&0&-e^{\imath k}\cr I_4&0&0&0\cr}$&
$\pmatrix{C^*_2&0&0&0\cr 0&C_2&0&0\cr 0&0&C_2^*&0\cr 0&0&0&C_2\cr}$\\ \hline
$_{\bf k}G^\pm$&8&$\pmatrix{B_4&0\cr 0&B_4^{-1}\cr}$&
$\pmatrix{0&I_4\cr I_4&0\cr}$&$\pm
\pmatrix{E_2&0&0&0\cr 0&E_2^*&0&0\cr 0&0&F_2&0\cr 0&0&0&F_2^*\cr}$&
$\pmatrix{K_2&0&0&0\cr 0&K_2^*&0&0\cr 0&0&K_2&0\cr 0&0&0&K_2^*\cr}$\\ 
\hline\hline
\end{tabular}}\end{center}\end{table}}}

\clearpage
\subsection{The irrs of the hexagonal groups}
The irrs of the hexagonal diperiodic groups ${\bf Dg}65$ - ${\bf Dg}80$
are presented in the tables \ref{65}-\ref{71} and the corresponding
irreducible domains of the {\Brillouin} zone are sketched in the Fig.
\ref{HEXid}. The primitive translations make the angle of
$\frac{2\pi}{3}$, while the angle between the basis vectors $k_1$ and
$k_2$ of the inverse lattice is $\pi/3$.

{\footnotesize
\begin{figure}[hbt]
\centereps{16cm}{8cm}{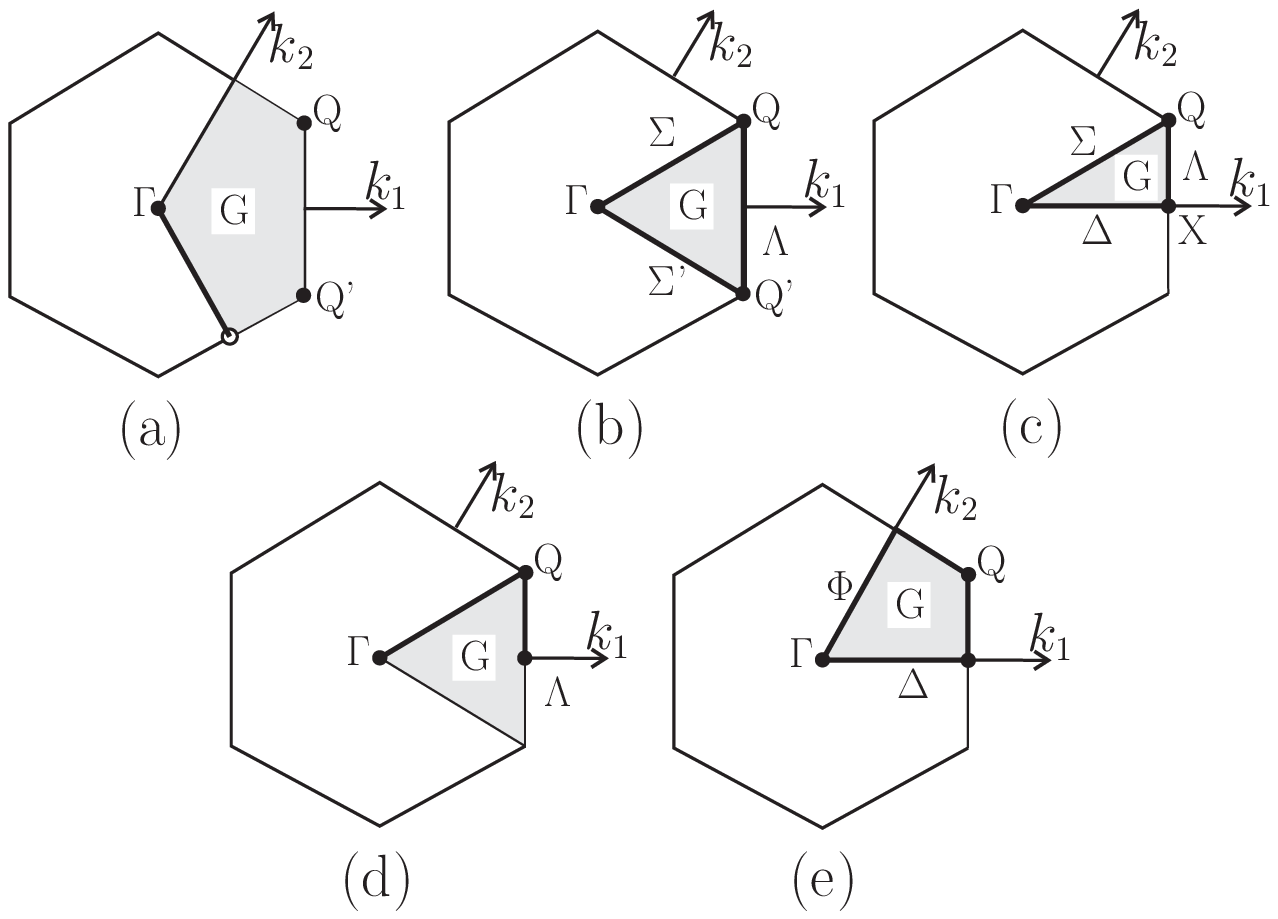}
\caption[]{\label{HEXid}{\footnotesize 
The irreducible domains of the {\Brillouin} zone for the hexagonal
diperiodic groups. The special points (filled circles) are $\gG
=(0,0)$, $X=(\pi ,0)$, $Q=(\frac{2\pi}{3},\frac{2\pi}{3})$ and
$Q'=(\frac{4\pi}{3},-\frac{2\pi}{3})$, while the special lines are $\gD
=(k,0)$ (for $k\in (0,\pi )$), $\Phi =(0,k)$ (for $k\in (0,\pi )$),
$\gS =(k,k)$ (for $k\in (0,\frac{2\pi}{3})$), $\gS '=(k,-\frac{k}{2})$
(for $k\in (0,\frac{4\pi}{3})$) and $\gL =(k,2\pi -2k)$ (where $k\in
(\frac{2\pi}{3},\pi )$ for the groups with the principle axis of the
order 6, and $k\in (\frac{2\pi}{3},\frac{4\pi}{3})$ for the groups
principle axis of the order 3).}}
\end{figure}}

The following abbreviations are used throughout the tables:\\
$B_3=\pmatrix{0&1\cr I_2&0\cr}$, 
$E_3={\rm diag}\, 
(e^{\imath k_1},e^{-\imath (k_1+k_2)},e^{\imath k_2})$, 
$J_3={\rm diag}\, (e^{\imath k},e^{\imath 2k},e^{\imath k})$,\\ 
$K_2={\rm diag}\, (e^{\imath 2\frac{\pi}{3}},e^{-\imath 2\frac{\pi}{3}})$, 
$K_3={\rm diag}\, (e^{\imath k},e^{-\imath k},e^{-2\imath k})$, 
$L_2={\rm diag}\, (e^{\imath k_2},e^{\imath k_1})$,\\ 
$L_3={\rm diag}\, (e^{-\imath 2k},e^{-\imath k},e^{\imath k})$, 
$O_3={\rm diag}\, (e^{\imath k_2},e^{\imath (k_1+k_2)},
e^{\imath k_1})$,  
$P_3={\rm diag}\, (e^{\imath k},1,e^{-\imath k})$,\\
$R_2=\pmatrix{0&e^{\imath 2\frac{\pi}{3}}\cr e^{\imath 2\frac{\pi}{3}}&0\cr}$, 
$R_3={\rm diag}\, (1,e^{\imath k},e^{\imath k})$ and 
$S_3={\rm diag}\, (e^{\imath k_1},e^{-\imath k_2},e^{-\imath
(k_1+k_2)})$. 
{\footnotesize{
\begin{table}[hbt]	
\caption[]{\label{65}{\footnotesize{
The irrs of the hexagonal diperiodic groups 
$\D\g 65={\bf C}_3\T=\gr\{C_3, \tx, \ty\}$ and 
$\D\g 74={\bf C}_{3h} \T =\gr\{C_3, \gs_h,\tx, \ty\}$,
induced by the elements of the axial point group ${\bf
C}_3$ from the irrs of the diperiodic groups $\D\g 1$ and $\D\g 4$
(Tab. \ref{1}), respectively.  The quasi angular momentum $m$ takes on
the values $-1,\, 0$ and $1$.  The irreducible domain of the
{\Brillouin} zone is presented in the Fig. \ref{HEXid}$\, (a)$.}}}
\begin{center}
\begin{tabular}{||c|c||c|c|c|c||} \hline \hline 
Irr&D& $C_3$& $\gs_h$& $(I|1\,0)$&$(I|0\,1)$ \\ \hline \hline
$\gG_m^{\pm}$& 1 & $e^{\imath \frac{2\pi}{3} m}$&$\pm 1$& 1 & 1  \\ \hline 
$Q_m^{\pm}$&1&$e^{\imath\frac{2\pi}{3}m}$&$\pm 1$&
$e^{\imath\frac{2\pi}{3}}$&$e^{\imath\frac{2\pi}{3}}$\\ \hline 
${Q'}_m^{\pm}$&1&$e^{\imath\frac{2\pi}{3}m}$&$\pm 1$&
$e^{-\imath\frac{2\pi}{3}}$&$e^{-\imath\frac{2\pi}{3}}$\\ \hline 
$_{\bf k}G^{\pm}$& 3 &$B_3$& $\pm I_3$&
$\pmatrix{e^{\imath k_1}&0&0\\0&e^{-\imath (k_1+k_2)}&0\\
0&0&e^{\imath k_2}\\}$
&$\pmatrix{e^{\imath k_2}&0&0\\0&e^{\imath k_1}&0\\
0&0&e^{-\imath (k_1+k_2)}\\}$ \\ \hline \hline 
\end{tabular}\end{center}\end{table}}}
{\footnotesize{
\begin{table}[hbt]
\caption[]{\label{68}{\footnotesize{
The irrs of the hexagonal diperiodic groups  
$\D\g 68={\bf D}_3\T =\gr\{C_3, U_x, \tx, \ty\}$, 
$\D\g 70={\bf C}_{3v}\T =\gr\{C_3, \gs_x, \tx, \ty\}$ and 
$\D\g 79={\bf D}_{3h}\T =\gr\{C_3, U_x, \gs_h, \tx, \ty\}$ induced by
the element $U_x$ or $\sigma_x$ from the irrs of the groups 
$\D\g 65$ and $\D\g 74$ (Tab. \ref{65}). 
The irreducible domain of the {\Brillouin} zone is presented in the 
Fig. \ref{HEXid}$\, (b)$.}}}
\begin{center}
\begin{tabular}{||c|c||c|c|c|c|c||} \hline\hline 
Irr&D&$C_3$&$\gs_x\ {\rm or}\ U_x$&$\gs_h$&$(I|1\,0)$&$(I|0\,1)$\\ \hline\hline
$^v\gG_0^\pm$&1&$1$&$(-1)^v$&$\pm 1$&1&1\\ \hline 
$^vQ_0^\pm$&1&$1$&$(-1)^v$&$\pm 1$&$e^{\imath\frac{2\pi}{3}}$&
$e^{\imath\frac{2\pi}{3}}$\\ \hline 
$^v{Q'}_0^\pm$&1&$1$&$(-1)^v$&$\pm 1$&$e^{-\imath\frac{2\pi}{3}}$&
$e^{-\imath\frac{2\pi}{3}}$\\ \hline 
$\gG_{1}^\pm$&2&$\pmatrix{e^{\imath\frac{2\pi}{3}}&0\cr
0&e^{-\imath\frac{2\pi}{3}}\cr}$&$\pmatrix{0&1\cr 1&0\cr}$&
$\pm I_2$&$I_2$&$I_2$\\ \hline 
$Q_{1}^\pm$&2&$\pmatrix{e^{\imath\frac{2\pi}{3}}&0\cr
0&e^{-\imath\frac{2\pi}{3}}\cr}$&$\pmatrix{0&1\cr 1&0\cr}$&
$\pm I_2$&$e^{\imath\frac{2\pi}{3}}I_2$&$e^{\imath\frac{2\pi}{3}}I_2$\\
\hline
${Q'}_{1}^\pm$&2&$\pmatrix{e^{\imath\frac{2\pi}{3}}&0\cr
0&e^{-\imath\frac{2\pi}{3}}\cr}$&$\pmatrix{0&1\cr 1&0\cr}$&
$\pm I_2$&$e^{-\imath\frac{2\pi}{3}}I_2$&$e^{-\imath\frac{2\pi}{3}}I_2$\\
\hline
$^v_k\Lambda^\pm$&3&$B_3$&$(-1)^v\pmatrix{A_2&0\cr 0&1\cr}$&$\pm I_3$&
$\pmatrix{e^{\imath k}I_2&0\cr 0&e^{-2\imath k}\cr}$&
$\pmatrix{e^{-2\imath k}&0\cr 0&e^{\imath k}I_2\cr}$\\ \hline
$^v_k\Sigma^\pm$&3&$B_3$&$(-1)^vA_3$&$\pm I_3$&
$\pmatrix{e^{\imath k}&0&0\cr 0&e^{-2\imath k}&0\cr 0&0&e^{\imath k}\cr}$&
$\pmatrix{e^{\imath k}I_2&0\cr 0&e^{-2\imath k}\cr}$\\ \hline
$^v_k\Sigma '^\pm$&3&$B_3$&$(-1)^v\pmatrix{1&0\cr 0&A_2}$&$\pm I_3$&
$\pmatrix{e^{\imath k}&0\cr 0&e^{-\imath\frac{k}{2}}I_2\cr}$&
$\pmatrix{e^{-\imath\frac{k}{2}}&0&0\cr 0&e^{\imath k}&0\cr
0&0&e^{-\imath\frac{k}{2}}\cr}$\\ \hline
$_{\bf k}G^\pm$&6&$\pmatrix{B_3&0\cr 0&B_3^{-1}}$&$
\pmatrix{0&I_3\cr I_3&0\cr}$&$\pm I_6$&$
\pmatrix{E_3&0\cr 0&E_3}$&$\pmatrix{L_2&0&0\cr 
0&e^{-\imath (k_1+k_2)}I_2&0\cr 0&0&L_2\cr}$ \\ \hline \hline 
\end{tabular}\end{center}\end{table}}}
{\footnotesize{
\begin{table}[hbt]
\caption[]{\label{67}{\footnotesize{
The irrs of the hexagonal diperiodic groups $\D\g 67={\bf
D}_3\T=\gr\{C_3, U, \tx, \ty\}$, $\D\g 69={\bf C}_{3v}\T =\gr\{C_3,
\gs,\tx, \ty\}$ and $\D\g 78={\bf D}_{3h}\T =\gr\{C_3, U, \gs_h,
\tx,\ty\}$ induced by the elements $U$ and $\sigma$ from the irrs of
the groups $\D\g 65$ and $\D\g 74$ (Tab. \ref{65}). The quasi angular
momentum $m$ takes on the values $-1$, $0$ and $1$.  The irreducible
domain of the {\Brillouin} zone is presented in the Fig. \ref{HEXid}$\,
(e)$. Here, $D_3={\rm diag}\, (e^{\imath (k_1+k_2)},e^{-\imath
k_2},e^{-\imath k_1})$ and $N_3={\rm diag}\, (e^{\imath k_2},e^{\imath
k_1},e^{-\imath (k_1+k_2)})$.}}}
\begin{center}
\begin{tabular}{||c|c||c|c|c|c|c||} \hline \hline 
Irr&D&$C_3$&$\gs\ {\rm or}\ U$&$\gs_h$&$(I|1\,0)$&$(I|0\,1)$\\ \hline\hline
$^v\gG_0^\pm$&1&$1$&$(-1)^v$&$\pm 1$&1&1\\ \hline 
$\gG_{1}^\pm$&2&$\pmatrix{e^{\imath\frac{2\pi}{3}}&0\cr
0&e^{-\imath\frac{2\pi}{3}}\cr}$&$A_2$&$\pm I_2$&$I_2$&$I_2$\\ \hline 
$Q_m^\pm$&2&$\pmatrix{e^{\imath\frac{2\pi}{3}m}&0\cr
0&e^{-\imath\frac{2\pi}{3}m}\cr}$&$A_2$&
$\pm I_2$&$\pmatrix{e^{\imath\frac{2\pi}{3}}&0\cr 
0&e^{-\imath\frac{2\pi}{3}}\cr}$&$\pmatrix{e^{\imath\frac{2\pi}{3}}&0\cr 
0&e^{-\imath\frac{2\pi}{3}}\cr}$\\ \hline
$^v_k\Delta^\pm$&3&$B_3$&$(-1)^v\pmatrix{1&0\cr 0&A_2}$&$\pm I_3$&
$\pmatrix{e^{\imath k}&0&0\cr 0&e^{-\imath k}&0\cr 0&0&1\cr}$&
$\pmatrix{1&0&0\cr 0&e^{\imath k}&0\cr 0&0&e^{-\imath k}\cr}$\\ \hline
$^v_k\Phi^\pm$&3&$B_3$&$(-1)^vA_3$&$\pm I_3$&
$\pmatrix{1&0&0\cr 0&e^{-\imath k}&0\cr 0&0&e^{\imath k}\cr}$&
$\pmatrix{e^{\imath k}&0&0\cr 0&1&0\cr 0&0&e^{-\imath k}\cr}$\\ \hline
$_{\bf k}G^\pm$&6&$\pmatrix{B_3&0\cr 0&B_3^{-1}}$&$
\pmatrix{0&I_3\cr I_3&0\cr}$&$\pm I_6$&$
\pmatrix{E_3&0\cr 0&D_3}$&$\pmatrix{N_3&0\cr 0&N^*_3\cr}$ \\ \hline \hline 
\end{tabular}\end{center}\end{table}}}
{\footnotesize{
\begin{table}[hbt]
\caption[]{\label{66}{\footnotesize{
The irrs of the hexagonal diperiodic groups  
$\D\g 66={\bf S}_6\T=\gr\{C_6\gs_h, \tx, \ty\}$,
$\D\g 73={\bf C}_6\T =\gr\{C_6, \tx, \ty\}$ and 
$\D\g 75={\bf C}_{6h}\T =\gr\{C_6, \gs_h, \tx, \ty\}$ 
induced by the elements of the axial point
groups ${\bf C}_6$ or ${\bf S}_6$ from the irrs of the groups 
$\D\g 1$ and $\D\g 4$ (Tab. \ref{1}). 
For the points $\gG$, $X$ and $Q$ the quasi angular 
momentum $m$ takes on the integer values from the intervals 
$[-2,3]$, $[0,1]$ and $[-1,1]$, respectively. 
The irreducible domain of the {\Brillouin} zone is presented in the 
Fig. \ref{HEXid}$\, (d)$.}}}
\begin{center}
\begin{tabular}{||c|c||c|c|c|c||} \hline \hline 
Irr&D&$C_6$ or $C_6\gs_h$&$\gs_h$&$(I|1\,0)$&$(I|0\,1)$\\ \hline\hline
$\gG_m^{\pm}$&1&$e^{\imath\frac{\pi}{3}m}$&$\pm 1$&1&1\\ \hline 
$Q_m^\pm$&2&$\pmatrix{0&e^{-\imath m\frac{2\pi}{3}}\cr 
e^{-\imath m\frac{2\pi}{3}}&0\cr}$&$\pm I_2$&
$\pmatrix{e^{\imath\frac{2\pi}{3}}&0\cr 0&e^{-\imath\frac{2\pi}{3}}\cr}$&
$\pmatrix{e^{\imath\frac{2\pi}{3}}&0\cr 0&e^{-\imath\frac{2\pi}{3}}\cr}$\\ 
\hline
$X^{\pm}_m$&3&$(-1)^m\, B_3^{-1}$&$\pm I_3$&
$\pmatrix{-I_2&0\cr 0&1\cr}$&$\pmatrix{1&0\cr 0&-I_2\cr}$\\ \hline
$_{\bf k}G^{\pm}$&6&$\pmatrix{0&1\\I_5&0\\}$&$\pm I_6$&
$\pmatrix{S_3&0\\0&S_3^{*}\\}$&$\pmatrix{O_3&0\\0&O_3^*\\}$\\ \hline \hline 
\end{tabular}\end{center}\end{table}}}
{\footnotesize{
\begin{table}[hbt]
\caption[]{\label{71}{\footnotesize{The irrs of the hexagonal diperiodic 
groups
$\D\g 71={\bf D}_{3d}\T =\gr\{C_6\gs_h, \gs_x, \tx, \ty\}$,
$\D\g 72={\bf D}_{3d}\T =\gr\{C_6\gs_h, U_x, \tx, \ty\}$, 
$\D\g 76={\bf D}_6\T=\gr\{C_6, U_x, \tx, \ty\}$,
 $\D\g 77={\bf C}_{6v}\T=\gr\{C_6, \gs_x, \tx, \ty\}$ and 
$\D\g 80={\bf D}_{6h}\T=\gr\{C_6, \gs_x,\gs_h, \tx, \ty\}$ 
induced by the elements $\gs_x$ or 
$U_x$ from the
irrs of the groups $\D\g 66$, $\D\g 73$ and $\D\g 75$ (Tab. \ref{66}).
For $^v\gG^{\pm}_m$, $\gG^{\pm}_{m}$ and 
$^vX^{\pm}_m$ the quasi angular momentum $m$ takes on the values from the
sets $\{ 0,3\}$, $\{ 1,2\}$ and $\{ 0,1\}$, respectively. 
The irreducible domain of the {\Brillouin}
zone is presented in the Fig. \ref{HEXid}$\, (c)$. Here, 
$V_3={\rm diag}\, (e^{-\imath (k_1+k_2)},e^{-\imath k_1},e^{\imath k_2})$.}}}
\begin{center}
\begin{tabular}{||c|c||c|c|c|c|c||} \hline
Irr&D&$C_6$ or $C_6\sigma_h$&$U_x$ or $\sigma_x$&$\sigma_h$&
$(I|1\,0)$&$(I|0\,1)$\\ \hline \hline
$^v\gG^{\pm}_m$&1&$(-1)^m$&$(-1)^v$&${\pm}1$&$1$&$1$\\ \hline
$\gG^{\pm}_{m}$&2&$\pmatrix{e^{\imath m\frac{\pi}{3}}&0\cr 
0&e^{-\imath m\frac{\pi}{3}}\cr}$&$\pmatrix{0&1\cr 1&0\cr}$&${\pm}I_2$&$I_2$&
$I_2$\\ \hline
$^vQ_0^\pm$&2&$\pmatrix{0&1\cr 1&0\cr}$&$(-1)^vI_2$&$\pm I_2$&$K_2$&$K_2$\\ 
\hline
$^vX_m^\pm$&3&$(-1)^m\, B_3^{-1}$&$(-1)^v\pmatrix{A_2&0\cr 0&1\cr}$&$\pm I_3$&
$\pmatrix{-I_2&0\cr 0&1\cr}$&$\pmatrix{1&0\cr 0&-I_2\cr}$\\ \hline
$Q_{1}^\pm$&4&$\pmatrix{R_2^*&0\cr 0&R_2\cr}$&
$\pmatrix{0&I_2\cr I_2&0\cr}$&$\pm I_4$&$\pmatrix{K_2&0\cr 0&K_2\cr}$&
$\pmatrix{K_2&0\cr 0&K_2\cr}$\\ \hline
$^v_k\Sigma^\pm$&6&$\pmatrix{0&1\cr I_5&0\cr}$&$(-1)^v
\pmatrix{A_5&0\cr 0&1\cr}$&$\pm I_6$&
$\pmatrix{K_3&0\cr 0&K_3^*\cr}$&$\pmatrix{J_3&0\cr 0&J_3^*\cr}$\\ \hline 
$^v_k\Delta^\pm$&6&$\pmatrix{0&1\cr I_5&0\cr}$&$(-1)^vA_6$&$\pm I_6$&
$\pmatrix{P_3&0\cr 0&P_3^*\cr}$&$\pmatrix{R_3&0\cr 0&R_3^*\cr}$\\ \hline 
$^v_k\Lambda^\pm$&6&$\pmatrix{0&1\cr I_5&0\cr}$&$(-1)^v
\pmatrix{A_3&0\cr 0&A_3\cr}$&$\pm I_6$&
$\pmatrix{J_3&0\cr 0&J_3^*\cr}$&$\pmatrix{L_3&0\cr 0&L_3^*\cr}$\\ \hline 
$_{\bf k}G^\pm$&12&$\pmatrix{0&1&0&0\cr I_5&0&0&0\cr 0&0&0&I_5\cr 
0&0&1&0\cr}$&$\pmatrix{0&I_6\cr I_6&0\cr}$&$\pm I_{12}$&
$\pmatrix{S_3&0&0&0\cr 0&S_3^*&0&0\cr 0&0&S_3&0\cr 0&0&0&S_3^*\cr}$&
$\pmatrix{O_3&0&0&0\cr 0&O_3^*&0&0\cr 0&0&V_3&0\cr 0&0&0&V_3^*\cr}$\\ 
\hline\hline
\end{tabular}\end{center}\end{table}}}

\clearpage
\section{Concluding remarks}
Using the factorization (table \ref{PRVA}) of the diperiodic groups
onto the generalized translational subgroup ${\bf Z}$, describing the
symmetry of the ordering of the elementary motifs, and the axial point
group {\bf P}, containing the symmetry of the single motif, the irrs
for all of the 80 diperiodic groups are found. They are presented by
the matrices of the generators of the groups (tables \ref{1}-\ref{71}).

These results enable the full application of symmetry in the studies of
the diperiodic physical systems. Although some rare attempts of the
symmetry analysis of such systems exist in the literature, they are
restricted to the isogonal point groups (IC and Raman spectra,
\cite{BURNEAU}, classification of the spin states, \cite{SIGRICE}),
or they use the irrs of the 3D space groups to derive the band
representations for a few relevant diperiodic groups,
\cite{DGph}. Also, the relevance of the diperiodic symmetry in the
context of the defects at surfaces and interfaces of 3D crystals is
notified, \cite{PO-HI}, but the applications involving the irreducible
representations were not available. It is possible now to make some
quite general physically relevant conclusions, which are to be
mentioned here.

The degeneracy of the quantum levels is equal to the dimension of
the corresponding irr. Since the dimension of the irrs is  at
most 12, this is the maximal orbital degeneracy possible in the
diperiodic systems; the other allowed degeneracies are 1, 2, 3, 4, 6
and 8.

The independent good quantum numbers are immediately given by the
symbol of the irr.  Thus, the diperiodical symmetry generally yields
good quantum numbers of quasi linear and quasi angular momenta, and
usually one or more types of parities (horizontal and vertical mirror
planes or rotations for $\pi$ around horizontal axes). The selection
rules related to these quantum numbers reflects the underlying
conservation laws; as usual, the linear quasi momenta are conserved or
changed for the vector of the inverse lattice (in the umklapp
processes), while the quantum number $m$ of the $z$-component of the
angular momenta is conserved up to the order of the principle axis. The
precise selection rules for any specific group can be calculated with
help of the Clebsch-Gordan coefficients.  Similarly, for the other
standard applications of the symmetry (independent components of the
physical tensors, phase transitions, classification of the vibrational
and electronic bands, etc.) the particular group must be considered;
still, some interesting examples will be briefly discussed here, as a
motivation for further research.

In the important paper \cite{HA-STO}, the phase transitions of the
diperiodic structures have been analysed within Landau theory. For that
purpose the real representations related to the special points are used
to find the Molien functions generating the invariants and the order
parameters. They are constructed by the subduction from the space
supergroup, and they essentially correspond to a part of our results.
In fact, the choice of the generators is not always the same (our is
based on the internal factorization of the diperiodic group, while in
\cite{HA-STO} it is determined by the correspondence of the invariants
of the diperiodic and space groups), while, due to the different
methods of construction the many dimensional representations do not
have the same matrices, although they are equivalent. In this context,
now it is possible to investigate the incommensurate transitions
(related to the special lines and general points) from the same point
of view.

While the first investigations on the diperiodic groups, \cite{EW},
reflected the interest for the semiconductors, the present paper has
been inspired by the theoretical investigations on the high temperature
superconducting materials. The most of those 3D structures are
periodical in two dimensions only, since such compounds contain the
different and even aperiodically (along the $z$-axis) arranged layers,
with several $CuO_2$ conducting layers always present.  The most
frequently discussed compounds are with the symmetry of ${\bf
Dg}61=\D_{4h}\T$, table \ref{53}. The possible topology of the band
shapes can be calculated by the compatibility relations: when the
general point ${\bf k}$ tends to the boundary of the irreducible
domain, i. e. to some of the special lines or points, the
representations $G^\pm$ become reducible, and their irreducible
components reveal the bands stickings at the boundary. This analysis
for the $CuO_2$ layers give the band shapes for the relevant electronic
states (Fig. \ref{Fez61}).
{\footnotesize
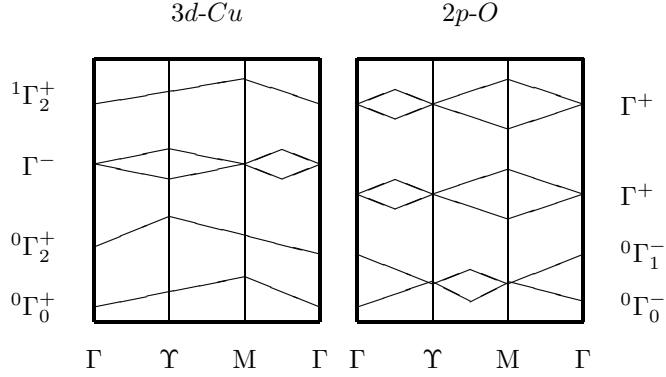
\begin{figure}[hbt]\begin{center}
\unitlength=1.00mm
\begin{picture}(90.00,50.00)
\linethickness{1pt}
\put(30.00,50.00){\makebox(0,0)[cb]{$3d$-$Cu$}}
\put(15.00,10.00){\line(1,0){30.00}}
\put(15.00,45.00){\line(1,0){30.00}}
\put(50.00,10.00){\line(1,0){30.00}}
\put(50.00,45.00){\line(1,0){30.00}}
\linethickness{0.4pt}
\put(25.00,10.00){\line(0,1){35.00}}
\put(35.00,10.00){\line(0,1){35.00}}
\put(15.00,5.00){\makebox(0,0)[cc]{$\gG$}}
\put(25.00,5.00){\makebox(0,0)[cc]{$\gY$}}
\put(35.00,5.00){\makebox(0,0)[cc]{$\M$}}
\put(45.00,5.00){\makebox(0,0)[cc]{$\gG$}}
\put(10.00,12.00){\makebox(0,0)[rc]{${^0\gG^+_0}$}}
\put(15.00,12.00){\line(5,1){20.00}}
\put(35.00,16.00){\line(5,-2){10.00}}
\put(10.00,20.00){\makebox(0,0)[rc]{${^0\gG^+_2}$}}
\put(15.00,20.00){\line(5,2){10.00}}
\put(25.00,24.00){\line(4,-1){20.00}}
\put(10.00,31.00){\makebox(0,0)[rc]{${\gG^-}$}}
\put(15.00,31.00){\line(5,1){10.00}}
\put(25.00,33.00){\line(5,-1){10.00}}
\put(35.00,31.00){\line(5,2){5.00}}
\put(40.00,33.00){\line(5,-2){5.00}}
\put(45.00,31.00){\line(-5,-2){5.00}}
\put(40.00,29.00){\line(-5,2){5.00}}
\put(35.00,31.00){\line(-5,-1){10.00}}
\put(25.00,29.00){\line(-5,1){10.00}}
\put(10.00,40.00){\makebox(0,0)[rc]{${^1\gG^+_2}$}}
\put(15.00,39.00){\line(6,1){20.00}}
\put(35.00,42.33){\line(3,-1){10.00}}
\linethickness{1pt}
\put(65.00,50.00){\makebox(0,0)[cb]{$2p$-$O$}}
\put(15.00,10.00){\line(0,1){35.00}}
\put(45.00,10.00){\line(0,1){35.00}}
\put(50.00,10.00){\line(0,1){35.00}}
\put(80.00,10.00){\line(0,1){35.00}}
\linethickness{0.4pt}
\put(60.00,10.00){\line(0,1){35.00}}
\put(70.00,10.00){\line(0,1){35.00}}
\put(50.00,5.00){\makebox(0,0)[cc]{$\gG$}}
\put(60.00,5.00){\makebox(0,0)[cc]{$\gY$}}
\put(70.00,5.00){\makebox(0,0)[cc]{$\M$}}
\put(80.00,5.00){\makebox(0,0)[cc]{$\gG$}}

\put(85.00,12.00){\makebox(0,0)[lc]{${^0\gG^-_0}$}}
\put(50.00,12.00){\line(3,1){10.00}}
\put(60.00,15.33){\line(2,-1){5.00}}
\put(65.00,12.67){\line(2,1){5.00}}
\put(70.00,15.33){\line(4,-1){10.00}}
\put(85.00,19.00){\makebox(0,0)[lc]{${^0\gG^-_1}$}}
\put(50.00,19.00){\line(5,-2){10.00}}
\put(60.00,15.00){\line(5,2){5.00}}
\put(65.00,17.00){\line(5,-2){5.00}}
\put(70.00,15.00){\line(5,2){10.00}}
\put(85.00,27.00){\makebox(0,0)[lc]{${\gG^+}$}}
\put(50.00,27.00){\line(5,2){5.00}}
\put(55.00,29.00){\line(5,-2){5.00}}
\put(60.00,27.00){\line(3,1){10.00}}
\put(70.00,30.33){\line(3,-1){10.00}}
\put(80.00,27.00){\line(-3,-1){10.00}}
\put(70.00,23.67){\line(-3,1){10.00}}
\put(60.00,27.00){\line(-5,-2){5.00}}
\put(55.00,25.00){\line(-5,2){5.00}}
\put(85.00,39.00){\makebox(0,0)[lc]{${\gG^+}$}}
\put(50.00,39.00){\line(5,2){5.00}}
\put(55.00,41.00){\line(5,-2){5.00}}
\put(60.00,39.00){\line(3,1){10.00}}
\put(70.00,42.33){\line(3,-1){10.00}}
\put(80.00,39.00){\line(-3,-1){10.00}}
\put(70.00,35.67){\line(-3,1){10.00}}
\put(60.00,39.00){\line(-5,-2){5.00}}
\put(55.00,37.00){\line(-5,2){5.00}}
\end{picture}
\caption[]{\label{Fez61} {\footnotesize{ The scheme of the bands
induced by $3d$-$Cu$ and $2p$-$O$ electronic orbitals of $CuO_2$ layer.
The energies are not calculated, and only the trends along the bands
are roughly estimated, \cite{ALTMAN2}, in order to represent the band
sticking together predicted by the compatibility relations of the irrs
$G^\pm$ at the boundary of the irreducible domain of the Brillouin
zone. The special positions at the boundary, specified below the
figures, are separated by the general positions of the interior.}}}
\end{center}\end{figure} }
Analogously, the symmetry classification of the ionic vibrations as
well as of the other electronic states of these superconducting
materials,  can be performed with help of the irrs of the diperiodic
groups. Although the detailed symmetry analysis of such systems will be
presented elsewhere, it should be stressed out that the simultaneous
classifications of the vibrational and electronic states has revealed
some kind of the anomalous vibronic coupling, \cite{G21-IY}:
surprisingly enough, it appears that there are degenerate electronic
states which are not coupled vibronically to the phonons. This is the
first known breakdown of the Jahn-Teller theorem, which had been
previously verified  for the molecules, \cite{JT}, polymers, \cite{IY},
and a number of the 3D crystals.  There are some experimental results
supporting this prediction, \cite{IVAN}.

Finally, let it be emphasized that the determination of generators of
the diperiodic groups (being simplified by the factorization of the
groups) enables the direct computer implementation. Indeed, the
modified group projector method, involving the only the generators, can
be employed straightforwardly. The program, analogous to the one
designed for the systems with the translational periodicity in one
direction, using this method, \cite{LESHOUCHES}, is in progress;
especially, it has been already applied to check the results of this
paper independently.

\appendix

\section*{Appendix A: Construction of irrs}
The structural properties of the diperiodic groups allow to avoid the
general induction procedure, and to reduce the task to the three
especially simple cases, being elaborated in the literature,
\cite{JABOON}.

{\em Method 1}: the group ${\bf G}$ is the product of its subgroups ${\bf H}$
and ${\bf K}$. The irrs of ${\bf H}$ and ${\bf K}$
($\{D^{(\mu)}({\bf H})\}$, and $\{D^{(\nu)}({\bf K})\}$) suffice to
find the irrs of ${\bf G}$: these are $D^{(\mu,\nu )} ({\bf G})$, defined by
$D^{(\mu,\nu)}(g)=D^{(\mu)}(h)\otimes D^{(\nu)}(k)$, for each element
$g=hk$ of ${\bf G}$, and each pair of irrs of ${\bf H}$ and ${\bf K}$.

{\em Method 2}: the group ${\bf G}$ is semidirect product of its subgroups
${\bf H}$ and ${\bf K}$, ${\bf H}$ being Abelian (with one-dimensional
irrs $\{\gD^{(\mu)}({\bf H})\}$). Then the subgroup ${\bf K_\mu}$ of
${\bf K}$ for each $\mu$, consists of the elements $l\in {\bf K}$
satisfying $\gD^{(\mu)}(l^{-1}hl)=Z^{-1}\gD^{(\mu)}(h)Z$ (for all $h$
in ${\bf H}$ and fixed nonsingular matrix $Z$).  For each irr
$\gd^{(\nu)}({\bf K}_\mu)$ of ${\bf K}_\mu$, the irr
$\gG^{(\mu,\nu)}(hk_\mu)=\gD^{(\mu)}(h)\otimes \gd^{(\nu)}(k_\mu)$ of
the little group ${\bf H}{\bf K_\mu}$ yields one induced irr of ${\bf
G}$: $D^{(\mu,\nu )}({\bf G})=\gG^{(\mu,\nu)}({\bf H}{\bf
K_\mu})\uparrow {\bf G}$.

In both cases the set of the generators of ${\bf G}$ is the union of the
sets of the  generators of  ${\bf H}$ and ${\bf K}$.

{\em Method 3}: the group is of the form $\G=\H+s\H$, where $\H$ is
halving subgroup (with the set of all nonequivalent irrs
$\{\gD^{(\mu)}(\H )\}$), and $s$ is an element of the coset of $\H$. If
there is nonsingular matrix $Z$ satisfying $Z^2=\gD^{(\mu)}(s^2)$ and
$\gD^{(\mu)}(s^{-1}hs)=Z^{-1}\gD^{(\mu)}(h)Z$, for each $h\in\H$, two
irrs of $\G$ are obtained: 
$$\{D^{(\mu\pm)}(h)\d=\gD^{(\mu)}(h),\
D^{(\mu\pm)}(sh)\d=\pm Z\gD^{(\mu)}(h)\,|\,h\in \H\};$$
if there is no such $Z$, then  the induced representation $D^{(\mu)}(\G
)=\gD^{(\mu)}(\H )\uparrow\G$, defined by the matrices
$$\{D^{(\mu)}(h)=\pmatrix{\gD^{(\mu)}(h)&0\cr
0&Z^{-1}\gD^{(\mu)}(h)Z\cr},\ D^{(\mu)}(sh)=
\pmatrix{0&\gD^{(\mu)}(s^2)Z^{-1}\gD^{(\mu)}(h)Z\cr \gD^{(\mu)}(h)&0\cr},$$
is irreducible. If $\{ h_1,\dots,h_k\}$ are the generators of $\H$,
then the set $\{h_1,\dots,h_k,s\}$ generates $\G$.  In some cases, this
set is not minimal, since some of the elements $h_i$ are monomials over
$s$ and the remaining generators.

To make the calculations completely transparent, the paths of the
induction will be given explicitly, for each type of the Brillouin
zone separately.  As for the oblique groups, ${\bf Dg}1$, ${\bf Dg}4$
and ${\bf Dg}5$ are direct products and their irrs are obtained by the
method 1. For the remaining groups the induction by the method 3 is
indicated by $\ind{g}$, where $g$ stands for the coset representative:
$${\bf Dg}1\cases{\ind{C_2\sigma_h}\D\g 2\cr \ind{C_2}\D\g
3\ind{(\sigma_h|0\,\frac{1}{2})}\D\g 7\cr},\quad
{\bf Dg}4\ind{C_2}\D\g 6.$$
The irrs of the rectangular groups are obtained by the method 3 in the
following 7 chains, each one starting from the one of the oblique
group. \\
\begin{minipage}{8cm}{
$${\bf Dg}1\cases{\ind{U_x}\D\g 8
             \cases{\ind{(I|\frac{1}{2}\,\frac{1}{2})}{\bf Dg}10\cr
                  \ind{(\sigma_h|\frac{1}{2}\,\frac{1}{2})}{\bf Dg}31\cr}\cr 
                  \ind{(U_y|0\,\frac{1}{2})}{\bf Dg}9
		\cases{\ind{C_2\sigma_h}\D\g 15\cr
		        \ind{C_2}\D\g 20\cr
			 \ind{\sigma_y}\D\g 26\cr}\cr
		    \ind{\sigma_x}\D\g 11
              \cases{\ind{(I|\frac{1}{2}\,\frac{1}{2})}{\bf Dg}13\cr
                  \ind{(\sigma_h|\frac{1}{2}\,\frac{1}{2})}{\bf Dg}32\cr}\cr
                  \ind{(\sigma_y|0\,\frac{1}{2})}{\bf Dg}12
   		\cases{\ind{C_2\sigma_h}\D\g 17\cr
		        \ind{U_x}\D\g 27\cr
			 \ind{C_2}\D\g 28\cr}\cr},$$
$${\bf Dg}3\cases{\ind{U_x}\D\g 19
             \cases{\ind{(I|\frac{1}{2}\,\frac{1}{2})}{\bf Dg}22\cr
                  \ind{(\sigma_h|\frac{1}{2}\,\frac{1}{2})}{\bf Dg}39\cr}\cr 
                  \ind{(U|\frac{1}{2}\,\frac{1}{2})}{\bf Dg}21\cr
		    \ind{\sigma_x}\D\g 23
              \cases{\ind{(I|\frac{1}{2}\,\frac{1}{2})}{\bf Dg}34\cr
                  \ind{(\sigma_h|\frac{1}{2}\,\frac{1}{2})}{\bf Dg}46\cr}\cr
                  \ind{(\sigma |\frac{1}{2}\,\frac{1}{2})}{\bf Dg}33\cr},$$}
\end{minipage}		    
\begin{minipage}{8cm}{
$${\bf Dg}2\cases{\ind{U_x}\D\g 14
             \cases{\ind{(I|\frac{1}{2}\,\frac{1}{2})}{\bf Dg}16\cr
                  \ind{(\sigma_h|\frac{1}{2}\,\frac{1}{2})}{\bf Dg}42\cr}\cr 
                  \ind{(\sigma |\frac{1}{2}\,\frac{1}{2})}{\bf Dg}18\cr},$$
$${\bf Dg}4\cases{\ind{U_x}\D\g 24
                  \ind{(I|\frac{1}{2}\,\frac{1}{2})}{\bf Dg}36\cr
                \ind{(U_y|0\,\frac{1}{2})}{\bf Dg}25\ind{C_2}{\bf Dg}40\cr},$$
$${\bf Dg}5\cases{\ind{U_x}\D\g 29\cr
                  \ind{(U_x|\frac{1}{2}\, 0)}{\bf Dg}30
              \cases{\ind{U_y}{\bf Dg}43\cr
         		\ind{C_2}{\bf Dg}45\cr}\cr
                  \ind{(I|\frac{1}{2}\,\frac{1}{2})}{\bf Dg}36\cr},$$
$${\bf Dg}6\cases{\ind{U_x}\D\g 37
                  \ind{(I|\frac{1}{2}\,\frac{1}{2})}{\bf Dg}47\cr
                  \ind{(\sigma |\frac{1}{2}\,\frac{1}{2})}{\bf Dg}44\cr},$$
$${\bf Dg}7\cases{\ind{U_y}\D\g 38\cr
                 \ind{\sigma_y}{\bf Dg}41
		   \ind{(I|\frac{1}{2}\,\frac{1}{2})}{\bf Dg}48\cr}.$$}
\end{minipage}\\
The irrs of the square groups are derived in two chains,
starting from the square oblique groups ${\bf Dg}1$ and ${\bf Dg}4$; besides
the induction from the halving subgroup, the method 2 has been used
(the subgroup $\K$ is in these cases indicated by $\Ind{\K}$):
$${\bf Dg}1\cases{\Ind{{\bf C}_4}\D\g 49
             \cases{\ind{(\sigma_h|\frac{1}{2}\,\frac{1}{2})}{\bf Dg}52\cr
                    \ind{U_x}\D\g 53\cr	      
                    \ind{(U|\frac{1}{2}\,\frac{1}{2})}{\bf Dg}54\cr
                    \ind{\sigma_x}\D\g 55\cr		      
                    \ind{(\sigma |\frac{1}{2}\,\frac{1}{2})}{\bf Dg}56\cr}\cr 
                  \Ind{{\bf S}_4}{\bf Dg}50
		\cases{\ind{U_x}\D\g 57\ind{(\sigma_h|\frac{1}{2}\,\frac{1}{2})}
				{\bf Dg}64\cr
		        \ind{(U|\frac{1}{2}\,\frac{1}{2})}\D\g 58\cr
			 \ind{U}\D\g 59\ind{(\sigma_h|\frac{1}{2}\,\frac{1}{2})}
				{\bf Dg}62\cr
                   \ind{(\sigma|\frac{1}{2}\,\frac{1}{2})}{\bf Dg}60\cr}\cr},\quad
{\bf Dg}4\Ind{{\bf C}_4}{\bf Dg}51\cases{\ind{U_x}\D\g 61\cr
                   \ind{(\sigma|\frac{1}{2}\,\frac{1}{2})}{\bf Dg}63\cr}.$$   
For the hexagonal groups there are two chains again, starting from the
hexagonal oblique groups ${\bf Dg}1$ and ${\bf Dg}4$, respectively.
Also, the second and the third induction procedures were necessary.
$${\bf Dg}1\cases{\Ind{{\bf C}_3}\D\g 65 \cases{\ind{U}{\bf Dg}67\cr
\ind{U_x}\D\g 68\cr \ind{\sigma}{\bf Dg}69\cr \ind{\sigma_x}\D\g
70\cr}\cr \Ind{{\bf S}_6}{\bf Dg}66 \cases{\ind{U}\D\g 71\cr
\ind{U_x}\D\g 72\cr}\cr \Ind{{\bf C}_6}{\bf Dg}73\ind{\sigma_x}{\bf
Dg}77\cr},\quad
{\bf Dg}4\cases{\Ind{{\bf C}_3}{\bf Dg}74\cases{\ind{U}\D\g 78\cr
\ind{U_x}{\bf Dg}79\cr}\cr \Ind{{\bf C}_6}{\bf Dg}75\ind{U_x}{\bf
Dg}80\cr}.$$


\end{document}